\documentclass[%
 aip,
 amsmath,amssymb,
 reprint,%
]{revtex4-1}

\usepackage{graphicx}
\usepackage{dcolumn}
\usepackage{bm}

\usepackage[utf8]{inputenc}
\usepackage[T1]{fontenc}
\usepackage{mathptmx}
\usepackage{etoolbox}

\usepackage{physics}
\usepackage{xcolor}
\usepackage{hyperref}
\usepackage[caption=false,justification=raggedright]{subfig}

\makeatletter
\def\@email#1#2{%
 \endgroup
 \patchcmd{\titleblock@produce}
  {\frontmatter@RRAPformat}
  {\frontmatter@RRAPformat{\produce@RRAP{*#1\href{mailto:#2}{#2}}}\frontmatter@RRAPformat}
  {}{}
}%
\makeatother
\begin{document}

\preprint{AIP/123-QED}

\title[]{Direct Comparison of Gyrokinetic and Fluid Scrape-Off Layer Simulations}

\author{A. Shukla} \affiliation{UT Austin}\affiliation{Exofusion}
\email{akashukla@utexas.edu.}
\author{J. Roeltgen} \affiliation{UT Austin}\affiliation{Exofusion}
\author{M. Kotschenreuther}\affiliation{Exofusion}
\author{J. Juno}\affiliation{Princeton Plasma Physics Laboratory}
\author{T. N. Bernard}\affiliation{General Atomics}
\author{A. Hakim}\affiliation{Princeton Plasma Physics Laboratory}
\author{G. W. Hammett}\affiliation{Princeton Plasma Physics Laboratory}
\author{D. R. Hatch}\affiliation{UT Austin}\affiliation{Exofusion}
\author{S. M. Mahajan}\affiliation{UT Austin}\affiliation{Exofusion}
\author{M. Francisquez}\affiliation{Princeton Plasma Physics Laboratory}

\date{\today}

\begin{abstract}
Typically, fluid simulations are used for tokamak divertor design. However, fluid models are only valid if the SOL is highly collisional, an assumption that is valid in many present day experiments but is questionable in the high-power scenarios envisioned for burning plasmas and fusion pilot plants. This paper reports on comparisons between fluid and kinetic simulations of the scrape off layer (SOL) for parameters and geometry representative of the Spherical Tokamak for Energy Production (STEP) fusion pilot plant. The SOLPS-ITER (fluid) and Gkeyll (gyroknietic) codes are operated in a two-dimensional (2D) axisymmetric mode, which replaces turbulence with ad-hoc diffusivities.
In kinetic simulations, we observe that the ions in the upstream SOL experience significant mirror trapping. This substantially increases the upstream temperature and has important implications for impurity dynamics. We show that the mirror force, which is excluded in SOLPS’s form of fluid equations, enhances the electrostatic potential drop along the field line in the SOL. We also show that the assumption of equal main ion and impurity temperatures, which is made in commonly used fluid codes, is invalid. 
The combination of these effects results in superior confinement of impurities to the divertor region in kinetic simulations, consistent with our earlier predictions~\cite{Mike23}. This effect can be dramatic, reducing the midplane impurity density by orders of magnitude. These results indicate that in reactor-like regimes the tolerable downstream impurity densities may be higher than would be predicted by fluid simulations, allowing for higher radiated power while avoiding unacceptable core contamination. Our results highlight the importance of kinetic simulations for divertor design and optimization for fusion pilot plants.
\end{abstract}

\maketitle


\section{Introduction}
Typically, fluid simulations are used to study the dynamics of the Scrape-Off Layer (SOL) in tokamaks~\cite{Hudoba2023,Osawa_2023,Rozhansky_2021, Zhang2024}.
These simulations are only valid if the SOL is highly collisional. However, fusion pilot plants will exist in a much less collisional parameter regime, and it is critical to understand the implications.  This paper reports on the investigation of SOL scenarios for the proposed Spherical Tokamak for Energy Production (STEP)~\cite{Baker2024}, systematically comparing fluid and gyrokinetic treatments of the parallel SOL physics in order to identify important kinetic effects.   We show that even in the nominal high recycling STEP scenario, the mean free path is long compared to the parallel temperature gradient length scale and kinetic effects have a large impact.  The simulations consist of 2D axisymmetric SOL simulations  using a fluid code, the B2.5 portion of the SOLPS-ITER~\citep{Schneider2006,SOLPS}
package, and a gyrokinetic code, Gkeyll~\citep{Mana23, Noah21, shi_thesis, Tess22, Gkeyll_website}. We find that the dynamics of the upstream SOL are kinetic—the velocity
distributions are not Maxwellian and there is significant mirror
trapping of the ions. This strongly increases the upstream ion temperature (roughly by a factor of three). We also find that in a magnetic configuration
with a Super-X like divertor, the mirror force accelerates
particles along the divertor leg resulting in an enhanced
potential drop along the field line~\cite{Mike23}. Finally, we demonstrate
that the assumption of equal ion and impurity temperatures often
made in fluid codes is violated. The combination of these
effects results in superior confinement of impurities to the divertor
region in kinetic simulations.

These findings have important implications for both high and low recycling regimes. 
Higher SOL ion temperature might lead to higher pedestal ion temperature. Via stiff transport, this can result in higher core fusion power. This could also increase the energy of neutrals hitting the main chamber wall, due to charge exchange of the hot plasma with cold recycled neutrals. This could increase wall erosion.
Also, the kinetic regimes characteristic of future devices may be able to support
larger downstream impurity densities (and hence more radiated power) than would be suggested by SOLPS. Higher confinement of impurities
to the divertor region would entail at least two benefits: (1) avoidance of impurity contamination of the core plasma, and (2) avoidance of high upstream densities, which can degrade confinement according to Ref.~\onlinecite{Mike23}.

We explore the differences in the steady state profiles and impurity distribution produced by both codes in several scenarios. We start with a base case comparison of highly collisional simulations in a slab to establish agreement between the two codes. We then investigate cases in the STEP outboard SOL geometry with and without impurities.

The article is outlined as follows: In section~\ref{sec:setup}, we describe the magnetic geometry, the simulation models used, and the simulations' setup. In section~\ref{sec:slab_comparison}, we establish baseline agreement between SOLPS and Gkeyll in simulations with a slab geometry. In section~\ref{sec:plasma_only}, we compare Gkeyll and SOLPS simulations in the outboard STEP SOL with no impurities. In section~\ref{sec:impurities}, we compare Gkeyll and SOLPS simulations in the outboard STEP SOL including argon impurities. Finally, we conclude in section~\ref{sec:conclusion}.

\section{Simulation Setup}\label{sec:setup}
Three types of simulations were employed in this work. First we ran simulations in a slab geometry with only electrons and deuterium ions to establish a baseline agreement between SOLPS and Gkeyll. Second, we ran simulations in the STEP geometry described in sec~\ref{sec:magnetic_geometry} with only electrons and deuterium ions. Finally, we added argon impurities to the STEP simulation and conducted simulations including charge states up to Ar$^{4+}$ and up to Ar$^{8+}$ with various neutral argon densities and temperatures. In this section we will describe the models used by both SOLPS and Gkeyll as well as the magnetic geometry, sources, and boundary conditions used in the simulations. Further details of the resolution and cost of each simulation as well as a link to the input files can be found in appendices~\ref{sec:resolution} and~\ref{sec:cost}.

\subsection{Model Descriptions}\label{sec:model_descriptions}
\subsubsection{Gkeyll}
\label{sec:Gkeyll_model}
Gkeyll is a full-f, long-wavelength gyrokinetic code using a Discontinuous Galerkin (DG) method for spatial discretization and Runge-Kutta for discretization in time. We use the electrostatic version of the code which solves the gyrokinetic equation
\begin{eqnarray}
\frac{\partial f_s}{\partial t}+\dot{\boldsymbol{R}} \cdot \nabla f_s+\dot{v}_{\|} \frac{\partial f_s}{\partial v_{\|}} - \nabla \cdot (\mathbf D \cdot \nabla f_s) = \nonumber \\
C\left[f_s\right]+S_s+C^{iz}_s+C^{rec}_s +  C^{rad}_s,
\label{eq:gkeq}
\end{eqnarray}
with $\dot{\mathbf{R}}=\{\mathbf{R}, H\}, \dot{v}_{\|}=\left\{v_{\|}, H\right\}, \textrm{ and } H_s=\frac{1}{2} m_s v_{\|}^2+\mu B+q_s \phi$
along with the gyrokinetic Poisson equation
\begin{equation}
-\nabla_{\perp} \cdot\left(\sum_s\frac{m_sn_{0s}}{B_0^2} \nabla_{\perp} \phi\right) = \sum_sq_s n_s(\boldsymbol{R}),
\label{eq:poisson}
\end{equation}
where  $f_s = f_s(\mathbf{R}, v_\parallel,\mu.t)$ is the gyrocenter distribution function for species $s$, $\mathbf R$ is the guiding center position, $\mathbf D$ is the particle diffusivity, $v_\parallel$ is the velocity parallel to the magnetic field, $\mu = \frac{mv_\perp^2}{2B}$ is the magnetic moment, $v_\perp$ is the velocity perpendicular to the magnetic field, $B$ is the magnetic field magnitude, $B_0$ is the magnetic field magnitude at the center of the simulation domain, $H_s$ is the gyrocenter center Hamiltonian of species s, $\phi$ is the electrostatic potential, $n_s$ is the guiding center density of species s, $n_{0s}$ is a reference density for species s, $q_i$ is the ion charge, and $e$ is the elementary charge.

The gyrokinetic Poisson bracket is given by~\citep{Noah21}
\begin{equation}
\{F, G\}=\frac{\boldsymbol{B}^*}{m B_{\|}^*} \cdot\left(\nabla F \frac{\partial G}{\partial v_{\|}}-\frac{\partial F}{\partial v_{\|}} \nabla G\right)-\frac{\mathbf{b}}{q B_{\|}^*} \times \nabla F \cdot \nabla G,
\end{equation}
with $\mathbf{B^*} = \mathbf{B} + (mv_\parallel/q)\nabla\times\mathbf{b}$ and $B_\parallel^* = \mathbf{b} \cdot \mathbf{B^*} \approx B$ where $\mathbf{b} = \mathbf{B}/B$ is the unit vector along the magnetic field.

The right-hand side of Eq.~\ref{eq:gkeq} contains the collision term~\citep{Mana22} $C[f_s]$, volumetric source terms $S_s$, ionization and recombination terms $C^{iz}_s$ and $C^{rec}_s$~\citep{Tess22}, and the radiation term $C^{rad}_s$~\cite{radiation}. 
The radiation term removes energy from the electrons based on the electron distribution function and the densities of the electrons and radiating impurities. Gkeyll's implementation of the radiation operator will be described in detail in a future publication associated with ref.~\onlinecite{radiation}.
Gkeyll uses the Dougherty collision operator~\cite{Dougherty1964, Mana22} which lacks the full velocity dependence of the full Fokker-Planck collision operator and thus does not match the Spitzer parallel heat conductivity in the collisional limit. Since we are comparing Gkeyll with SOLPS, we have adjusted the collision frequency in Gkeyll such that the parallel heat conductivity matches the Spitzer conductivity in the collisional limit. This amounts to a factor of 4 reduction in the collision frequency in Gkeyll. 

Note that Gkeyll implements the drift-kinetic limit of the gyrokinetic equation, neglecting all gyroaveraging operations. However, first order finite Larmor radius (FLR) effects are present in the ion polarization term in the gyrokinetic poisson equation, which distinguishes the long-wavelength gyrokinetic model from a drift-kinetic model~\cite{Noah21}.

In this paper, we conduct 2D axisymmetric simulations with field aligned coordinates, radial coordinate $\psi$ and poloidal coordinate $\theta$, as described in section~\ref{sec:magnetic_geometry}; our simulation grid does not include the binormal direction. So, like SOLPS, we are not directly simulating the turbulence and instead account for it by including radial diffusion as described in section~\ref{sec:sources}. 
We have also turned off drifts (the $\mathbf{E}\times \mathbf{B}$ drift, the $\nabla \mathbf{B}$ drift, and curvature drift) in both codes to simplify our comparison. In Gkeyll's axisymmetric model, turning off drifts amounts to setting the binormal component of the magnetic field unit vector $\mathbf b$ to zero.
Gkeyll's axisymmetric model does not yet support dynamic neutral species. So, for simulations with impurities, we use a static background neutral species as described in section~\ref{sec:impurity_sources}.


\subsubsection{SOLPS}
The SOLPS-ITER package consists of a 2D multi-fluid transport code, B2.5 and a 3D kinetic Monte Carlo neutral transport code EIRENE. We use only the fluid plasma edge solver B2.5 part of the SOLPS-ITER package described in Refs.~\onlinecite{Schneider2006} and~\onlinecite{SOLPS} in our simulations, since we are not evolving the neutral species. 
B2.5 evolves a continuity and momentum equation for each species, a total ion energy equation (one energy equation for all the ion species), and an electron energy equation. 
Section~\ref{sec:impurity_sources} describes how we freeze the neutral species in B2.5. Details of the SOLPS model equations can be found in Ref.~\onlinecite{Schneider2006}.

\subsubsection{Differences between Models}
There are 3 main differences between the two models. First, SOLPS makes the fluid assumption of lowest order Maxwellian velocity distributions for all species while Gkeyll does not. Second, SOLPS has just one energy equation for all of the ion species, which means that all ion species will have a common temperature; Gkeyll does not group the ions together so they are free to evolve to have different temperatures if the physics so dictates. Third, the mirror force is included via the parallel acceleration term in Gkeyll but not in SOLPS.
Note that while SOLPS's form of the fluid equations does not include the mirror force, it is possible to include some effects of the mirror force in a fluid code by including temperature anisotropy as is done in the 2D transport code UEDGE~\cite{Zhao_2021, Zhao2021_2, Zhao2022}. 

\subsection{Magnetic Geometry}\label{sec:magnetic_geometry}
\begin{figure}
    \centering
    \subfloat[Simulation domain used for STEP SOL simulations. The solid black lines indicate two flux surfaces, the separatrix and the outermost flux surface of the domain. The blue lines indicate the inner and outer radial boundaries of the simulation domain and the orange lines indicate the divertor plates which are the parallel boundaries. The simulation domain shown is 6cm wide at the OMP and the connection length from midplane to divertor plate is approximately 60m.]{
    \includegraphics[width=0.34\textwidth]{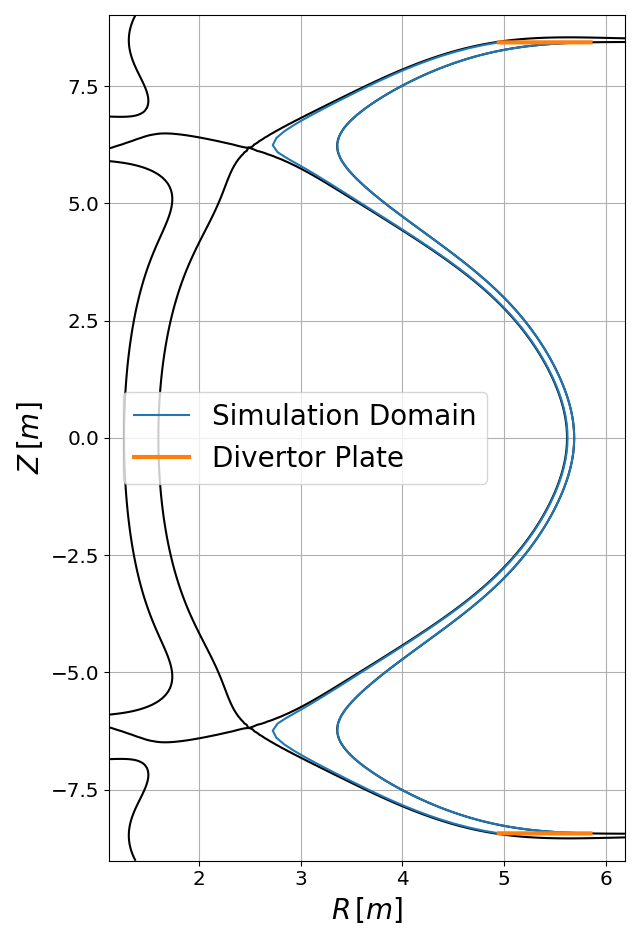}
    }\\
    \subfloat[Simulation domain used for slab simulations. This is a straight SOL which is 6cm wide and 120m long to match the width and connection length of the STEP simulation domain.]{
    \includegraphics[width=0.34\textwidth]{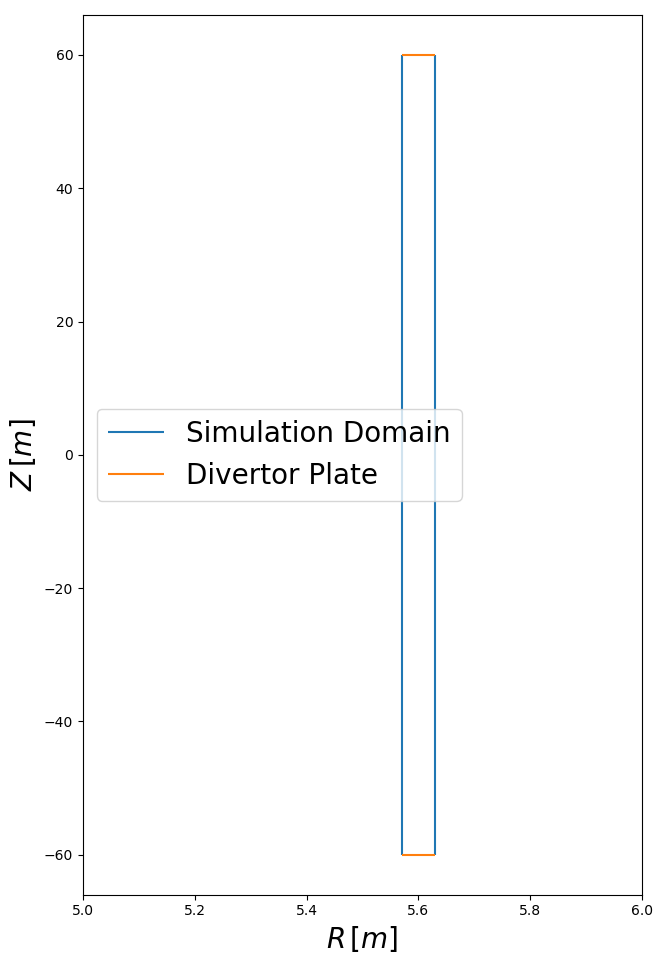}
    }
    \caption{
    STEP (a) and slab (b) simulation domains.
    \label{fig:geometry}
    }
\end{figure}

For simulations in the STEP geometry we use the equilibrium depicted in Fig.~\ref{fig:geometry}. This is a prospective equilibrium for STEP similar to the ones described in Ref.~\onlinecite{Hudoba2023}. The parameters of the equilibrium are listed in Table~\ref{tab:params}.
\begin{table}
  \begin{center}
\def~{\hphantom{0}}
  \begin{tabular}{|l|c|}
      \hline
      $R$        &  4 m    \\
      \hline
      $R_{axis}$ & 4.24 m \\
      \hline
      $a$        &  2 m    \\
      \hline
      $\kappa$   &  3.1    \\
      \hline
      $A$        &  2      \\
      \hline
      $I_p$      &  20.37 MA  \\
      \hline
      $B_{0}$ & 2.88 T     \\
      \hline
  \end{tabular}
  \caption{Magnetic Equilibirum Parameters: Major radius $R$, magnetic axis $R_{axis}$, minor radius $a$, elongation $\kappa$, aspect ratio $A$, plasma current $I_p$, toroidal magnetic field at Major radius, $B_0$.
  \label{tab:params}
  }
  \end{center}
\end{table}

The connection length from midplane to divertor plate is approximately 60m. We are targeting an SOL width of 2mm at the outboard midplane (OMP), so we chose to make the simulation domain much wider than that--3cm in Gkeyll and 6cm in SOLPS. We found that this width was enough to decouple the effect of the radial boundary conditions described in section~\ref{sec:BCs} from the dynamics in the main heat channel in both codes.

Gkeyll uses a field aligned coordinate system with poloidal flux ($\psi$) as the radial coordinate and normalized poloidal arc length ($\theta$) as the parallel coordinate. The parallel coordinate is normalized such that the upper and lower divertors are located at $\theta= \pm \pi$ respectively. The poloidal flux at the separatrix is $\psi_{sep} = 1.50982$ and $\psi$ increases toward the magnetic axis. Throughout this paper, we plot profiles using Gkeyll's coordinate system ($\psi, \theta$).

Gkeyll and SOLPS STEP simulations are both radially centered at the flux surface $\psi=1.2014$ which is 3.7cm away from the separatrix at the outboard midplane. The SOLPS simulation domain extends from $\psi = 0.934$ to $\psi = 1.469$ and the Gkeyll simulation domain extends from $\psi = 1.068$ to $\psi = 1.335$.

For the simulation in slab geometry, we use a box which is 6cm wide in the radial direction and 120m long in the direction parallel to the magnetic field to mimic the STEP SOL simulations.

\subsection{Sources \& Diffusion}\label{sec:sources}
For simulations in both the slab and STEP geometry, we use a particle source which is a Gausian centered at the radial center of the domain with a 0.76mm half-width at the OMP. The particle source is also a Gaussian along the field line constructed such that its value at the X-point is $10^2$ times smaller than its peak value at the OMP. Since the slab simulation does not have X-points, we construct the source so that it decays by a factor of $10^5$ 40m from the OMP. This decay length is approximately the same as in the STEP geometry.

\begin{figure}
    \subfloat[]{
    \includegraphics[width=0.45\textwidth]{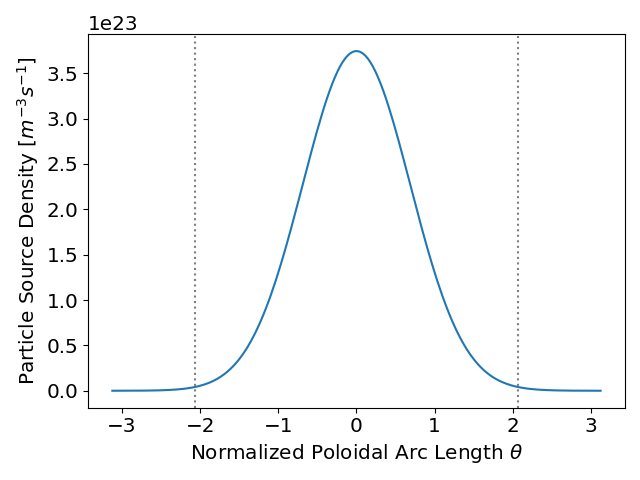}
    }\\
    \subfloat[]{
    \includegraphics[width=0.45\textwidth]{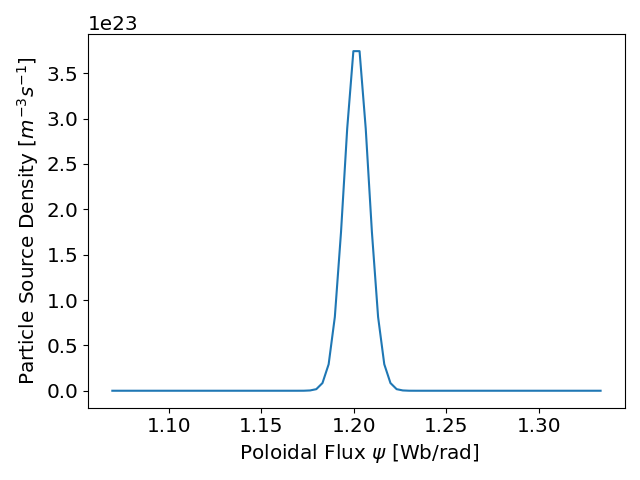}
    }
    \caption{
    Particle source density plotted along the field line at the radial center of the domain (a), and radially at the midplane (b). The source is constructed such that the value at the X-points, located at  $\theta = \pm$ 2.07 (marked by the gray dotted vertical lines), is 10$^2$ times smaller than the peak value. The radial half-width of the source at the OMP is 0.76mm.
    \label{fig:sourcedensity_plasma}
    }
\end{figure}

The source is localized between the X-points to mimic leakage of particles from the core into the SOL. We chose to center it at the radial center of the domain rather than the inner radial edge to decouple the effect of the boundary conditions from the dynamics in the main heat channel. In the radial direction, we chose the minimum width that could be represented without introducing numerical issues. We chose the minimum width because, in reality, a particle source in the SOL has no width; the particle source is radial diffusion of particles from the core.
The source density profiles, which are common to SOLPS and Gkeyll, are plotted radially and along the field line in Fig.~\ref{fig:sourcedensity_plasma} and are given by
\begin{equation}
n_{source}(R,Z) = \dot{n}_{peak}exp\{ -(R-R_{center})^2/2c_R^2 \} exp\{ -Z^2/2c_Z^2 \}
\end{equation}
with $\dot{n}_{peak} = 6.1 \times 10^{23} m^{-3}s^{-1}$, $c_R = 0.00074$ m, $c_Z = 7.22$ m, and $R_{center} = 0.003$ m for the slab simulations and
\begin{equation}
n_{source}(\psi,\theta) = \dot{n}_{peak}exp\{ -(\psi-\psi_{center}^2/2c_\psi^2 \}exp\{ -\theta^2/2c_\theta^2 \}
\end{equation}
with $\dot{n}_{peak} = 3.9 \times 10^{23} m^{-3}s^{-1}$, $c_\psi = 0.0065$ Wb/rad, $c_\theta = 0.688$, and $\psi_{center} = 1.2014$ Wb/rad for the STEP simulations.

We were targeting upstream parameters typical of a conventional high recycling regime, an upstream density of $3 \times 10^{19}$ and 80MW of input power, so we adjusted the particle source amplitude and temperature accordingly. The temperature of the source is 1.037 keV for both electrons and deuterium. The choice of 80 MW of input power is based on 100 MW of input power crossing the separatrix~\citep{Hudoba2023} with 20 MW going to the inboard SOL and 80 MW going to the outboard.

We ran both Gkeyll and SOLPS without drifts and with constant radial diffusion coefficients. The particle diffusivity was $D_\perp = 0.03$ and heat diffusivity was $\chi_\perp = 0.045$. These diffusivities were chosen in combination with the particle source width so that the half width of heat flux channel was approximately 2mm in the steady state. The proportion $\chi_\perp/D_\perp = 3/2$ is fixed by Gkeyll's implementation of diffusion; Gkeyll's diffusion coefficient, $\mathbf D$, introduced in section~\ref{sec:Gkeyll_model} is applied to the distribution function and does not depend on velocity.

\subsection{Sources of Impurities}
\label{sec:impurity_sources}
\begin{figure}
    \includegraphics[width=\columnwidth]{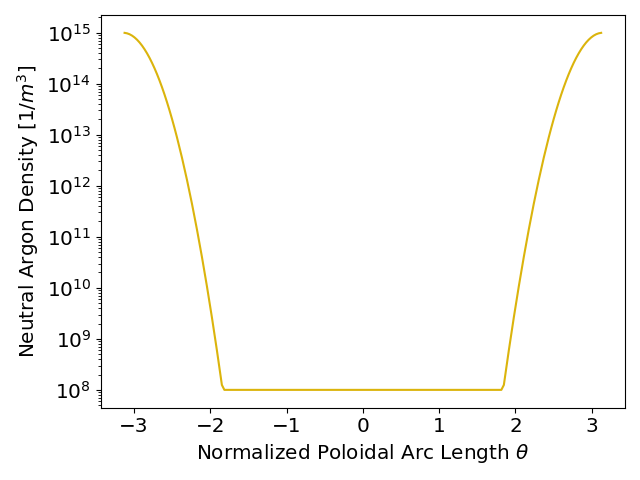}
    \caption{Neutral argon profile along the field line. The profile is uniform in $\psi$. The shape of the profile, peaked near the divertor plates, mimics gas puffing in the target region. In all simulations, we use the same shape for the profile and scale it to have a specified density at the divertor plate. The minimum density for the profile is always $n = 10^8 \textrm{ m}^{-3}$.
    \label{fig:Ar0}
    }
\end{figure}

In simulations with impurities we use a static background neutral argon profile concentrated downstream near the divertor plates as shown in Fig.~\ref{fig:Ar0}. The neutral argon profile is uniform in the radial direction ($\psi$) and is given by
\begin{equation}
n(\theta) =
\begin{cases}
    max(n_{peak} exp\{ -(\theta-\pi)^2/2c_\theta^2 \}, 10^8), & \theta \geq 0 \\
    max(n_{peak} exp\{ -(\theta+\pi)^2/2c_\theta^2 \}, 10^8), & \theta < 0 \\
\end{cases}
\end{equation}
with $n_{peak}$ varying between simulations and $c_{\theta} = 0.25$.
We include ionization and recombination for the neutral and charged argon species.

Gkeyll explicitly supports static neutral species while SOLPS does not. In SOLPS, we approximate a static neutral species by turning off the continuity and momentum equations for the fluid neutrals. The fluid neutrals still enter into the ion energy equation, but we expect the effect on the ion energy equation to be minimal due to the comparatively low neutral density.

We acknowledge that including dynamic neutrals in Gkeyll and using the full SOLPS-ITER package (B2.5 coupled to EIRENE) would be preferable to running these codes with static neutrals. However, Gkeyll's axisymmetric model does not currently support dynamic neutrals. We believe that, qualitatively, the major results of this work will hold even when dynamic  neutrals are included. In a future work, we will conduct a comparison including dynamic neutrals.

\subsection{Boundary Conditions}
\label{sec:BCs}
We adjusted the boundary conditions in the two codes to be as similar as possible. In the parallel (poloidal) direction, we use sheath boundary conditions on the particles. 
In Gkeyll, there is no boundary condition on the potential in the parallel direction~\citep{shi_thesis}. In Gkeyll's conducting sheath boundary condition, the potential at the sheath entrance, $\phi_{sh}$, is obtained from the gyrokinetic Poisson equation, Eq.~\ref{eq:poisson}, and electrons with energy less than $e\phi_{sh}$ are reflected back into the simulation domain~\cite{shi_thesis, Shi17}. 
In SOLPS STEP simulations, we use the standard sheath boundary condition which enforces the Bohm sheath condition at the sheath entrance~\cite{Schneider2006}.
In SOLPS slab simulations, we used an alternative sheath boundary condition for the energy equation that allows one to specify the sheath heat transmission coefficient for each species.
This is option 12 in the SOLPS manual~\cite{Schneider2006} and enforces a heat flux to the sheath entrance of $Q_s = \delta_s \Gamma_s T_s$ where $Q_s$ is the heat flux of species s, $\delta_s$ is the transmission coefficient for  species s, $T_s$ is the temperature of species s, and $\Gamma_s$ is the particle flux of species s. We used this boundary condition to match the electron and ion transmission coefficients in SOLPS to the radially averaged transmission coefficients observed in Gkeyll which were 4.62 and 2.88 respectively.

In Gkeyll, we apply absorbing boundary conditions to the particles in the radial direction--all particle species leave freely at the radial boundaries. We apply Dirichlet boundary conditions, $\phi=0$, to the potential at both radial boundaries for both Gkeyll and SOLPS.
For SOLPS, we use leakage boundary conditions in the radial direction with a large leakage coefficient of -1000 for both the continuity and energy equations.
The leakage boundary condition allows one to specify the radial velocity at the boundary as the leakage coefficient times the species' thermal velocity and the radial particle flux to the boundary as a leakage coefficient times the density times the sound speed. This leakage boundary condition with a large coefficient mimics the absorbing radial boundary conditions used in Gkeyll.
For the radial boundary condition of the momentum equation, we set the radial derivative of the momentum to zero to minimize the effect of the boundary condition on the momentum. Using a leakage boundary condition in the momentum equation would lead to an unphysically large velocity.

\section{Slab Comparison}\label{sec:slab_comparison}
As a base case comparison to establish reasonable agreement, we set up a simulation in a 2D (radial, poloidal) slab geometry without drifts. For this simulation we reduced the input power from 80MW to 50MW to target a more collisional regime where the codes are likely to agree. 
These simulations serve to verify that we are in fact using equivalent boundary conditions, sourcing, and diffusion in the two codes.

\begin{figure}
  \centering
  \subfloat[]{\label{fig:slab_density_x}
    \includegraphics[width=0.34\textwidth]{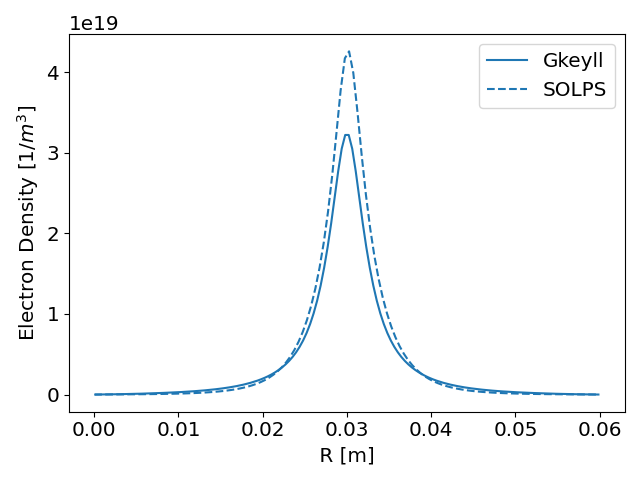}
  }\\
  \subfloat[]{\label{fig:slab_density_z}
    \includegraphics[width=0.34\textwidth]{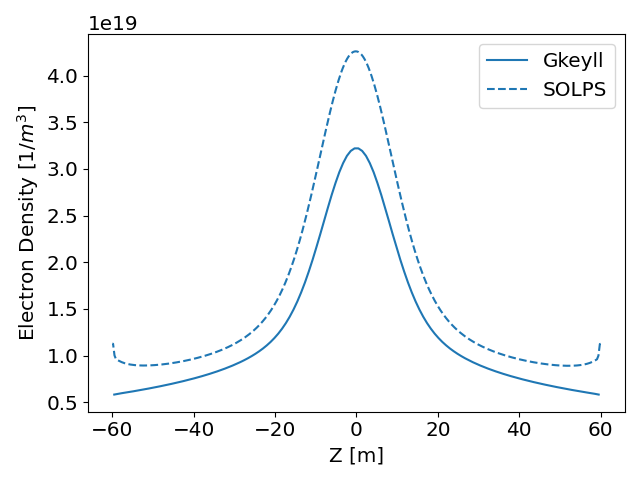}
  }\\
  \subfloat[]{\label{fig:slab_temp_x}
    \includegraphics[width=0.34\textwidth]{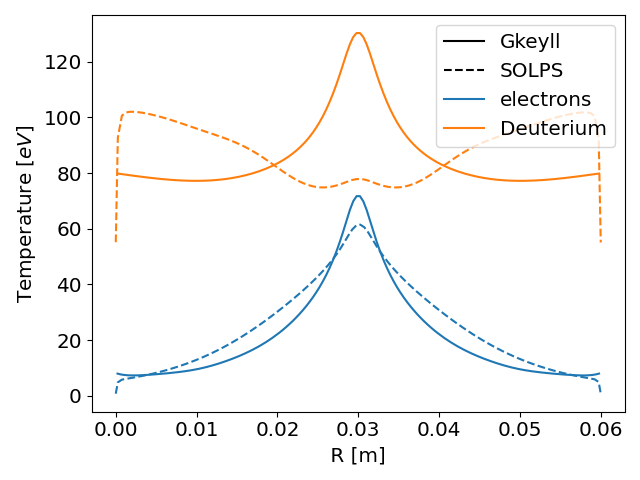}
  }\\
  \subfloat[]{\label{fig:slab_temp_z}
    \includegraphics[width=0.34\textwidth]{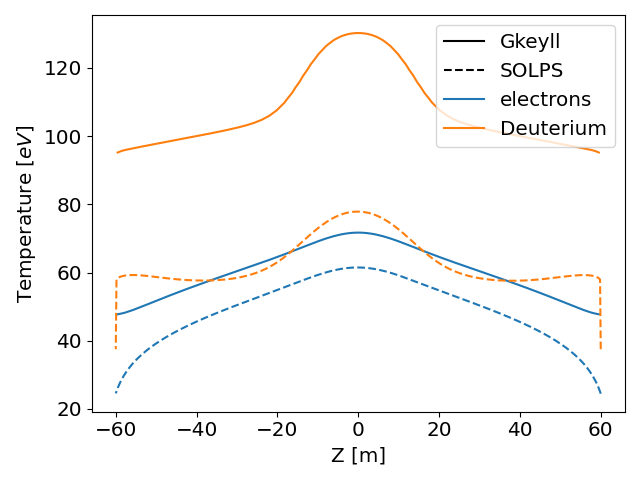}
  }
  \caption{
  Electron density plotted (a) radially (vs. R) and (b) along the field line (vs. Z)  and electron and ion temperature plotted (c) radially and (d) along the field line for both SOLPS and Gkeyll from the simulations with slab geometry. The density profiles are similar, but kinetic effects influence the parallel heat conduction and raise the ion temperature in Gkeyll relative to SOLPS.
    \label{fig:slab_moments}
    }
\end{figure}

\begin{figure}
    \subfloat[\label{fig:mfp_norm}]{
        \includegraphics[width=0.45\textwidth]{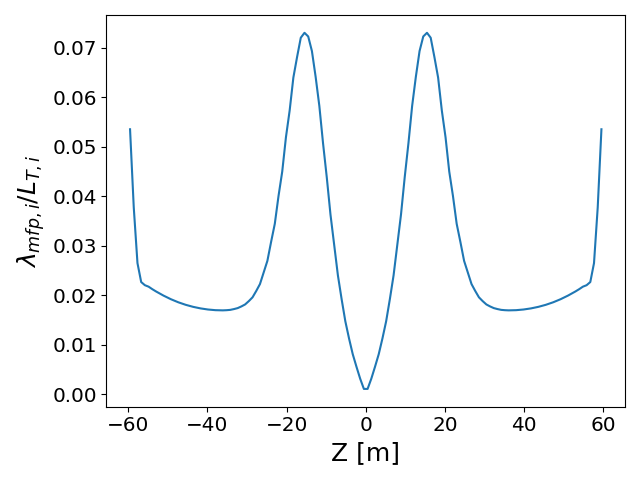}
    }\\
    \subfloat[\label{fig:mfp_temp}]{
        \includegraphics[width=0.45\textwidth]{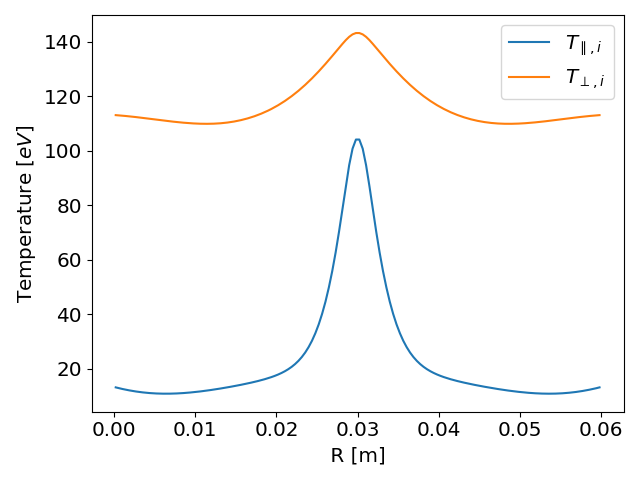}
    }
    \caption{
    Ion mean free  path normalized to $L_T$ from the Gkeyll simulation with slab geometry plotted along the field line at the radial center (a). Parallel temperature, $T_\parallel$, and perpendicular temperature, $T_\perp$, plotted radially at the midplane (b). The ion mean free path is not short enough for the ions to remain Maxwellian. The difference between the parallel and perpendicular temperatures in (b) demonstrates this.
    \label{fig:mfp_slab}
    }
\end{figure}

\begin{figure}
    \subfloat[]{
        \includegraphics[width=0.45\textwidth]{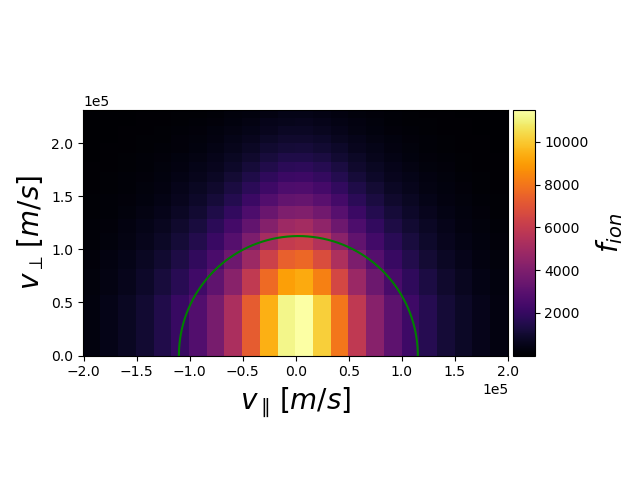}
    }\\
    \subfloat[]{
        \includegraphics[width=0.45\textwidth]{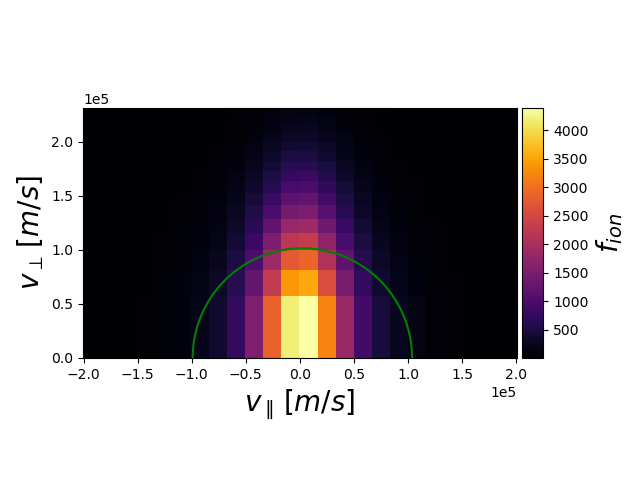}
    }
    \caption{
    Ion distribution function at midplane at R = 0.3cm (a) and R = 0.34cm (b) from Gkeyll simulations in slab geometry. The green line is a contour of a Maxwellian with the same temperature as the distribution function. The ion distribution function is not Maxwellian; it is depleted at high $v_\parallel$ because collisions cannot replenish the distribution function fast enough. This is true at both radial locations, but more obvious away from the radial center of the heat channel at R = 0.34 cm.
    \label{fig:slab_dist}
    }
\end{figure}

In Fig.~\ref{fig:slab_moments}, we plot the steady state temperature and density profiles along the field line and radially. The density and electron temperature profiles in SOLPS and Gkeyll agree quite well, but there is a difference between the ion temperatures. The reason for this difference is that, with 50 MW of input power, the ion mean free path, $\lambda_{mfp}$, is still not small enough that the ions are maxwellian. We can see this in Fig.~\ref{fig:mfp_norm} where we show the ion mean free path normalized to the parallel length scale, $L_{\parallel} = L_T$, where $L_T$ is the parallel temperature gradient scale length. Thus, kinetic effects influence the parallel heat conduction and raise the ion temperature in Gkeyll relative to SOLPS.
This is consistent with previous work~\cite{Xianzhu14} in which it has been shown that the conventional Braginskii fluid closure for the heat flux is only recovered if $\lambda_{mfp}/L_{\parallel}\lesssim5\times10^{-3}$.
As seen in fig Fig.~\ref{fig:slab_dist}, the ion distribution is depleted at large $v_\parallel$ because higher $v_\parallel$ particles leave faster and collisions cannot replenish the distribution function fast enough. This depletion at large $v_\parallel$ means that heat leaves more slowly along the field line and the ion temperature is raised. This results in $T_\parallel < T_\perp$ as shown in Fig.~\ref{fig:mfp_temp}.

In a very collisional regime where $\lambda_{mfp} \ll L_\parallel$, the ion temperatures would likely agree better, but we found that it was not feasible to run Gkeyll in this regime. As the collision frequency becomes very large, reducing the maximum stable time step, it becomes difficult to run a kinetic code with an explicit collision operator like Gkeyll. We are planning to utilize implicit collision operators in the future to avoid this limitation.

\section{Plasma Only Simulations}\label{sec:plasma_only}
As a next step, we ran simulations in the STEP SOL geometry. Since we kept drifts turned off for both codes and matched the diffusivities, differences between the two simulations beyond what was seen in section~\ref{sec:slab_comparison} can be attributed to the mirror force and non-Maxwellian velocity distributions.

\begin{figure}
    \subfloat[]{
        \includegraphics[width=0.45\textwidth]{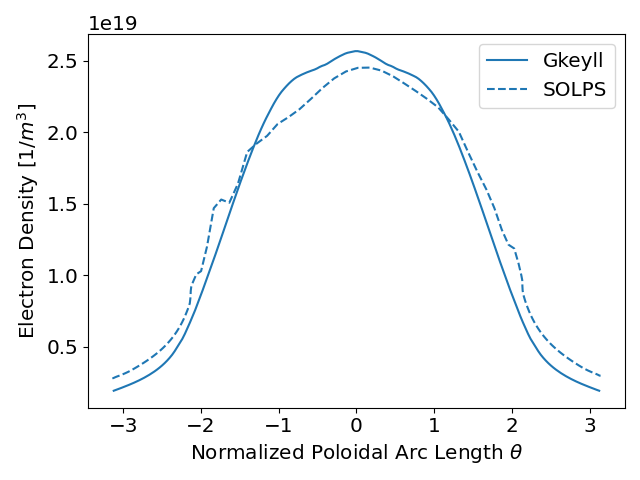}
    }\\
    \subfloat[]{
        \includegraphics[width=0.45\textwidth]{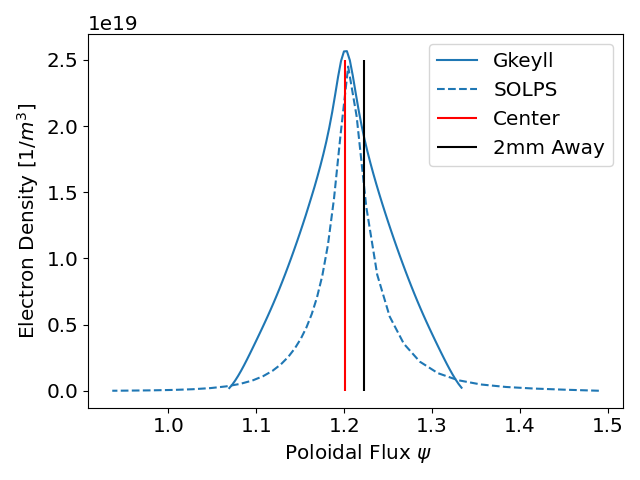}
    }
    \caption{
    Electron density plotted along the field line at the radial center of the domain (a), and radially at the midplane (b) for Gkeyll (solid) and SOLPS (dashed) simulations in STEP geometry. The vertical red line marks the flux surface at the radial center of the simulation at the OMP and the black line marks a surface 2mm away from the radial center. The density profiles given by Gkeyll and SOLPS are similar along  the field, but Gkeyll's radial density profile is slightly wider due to mirror trapping of the ions which broadens the potential and in turn broadens the electron density.
    \label{fig:density_plasma}
    }
\end{figure}

\begin{figure}
    \subfloat[\label{fig:temperature_plasma_z}]{
        \includegraphics[width=0.45\textwidth]{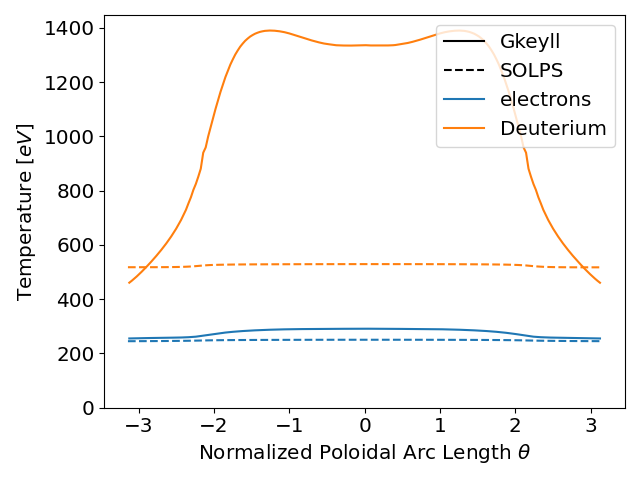}
    }\\
    \subfloat[]{
        \includegraphics[width=0.45\textwidth]{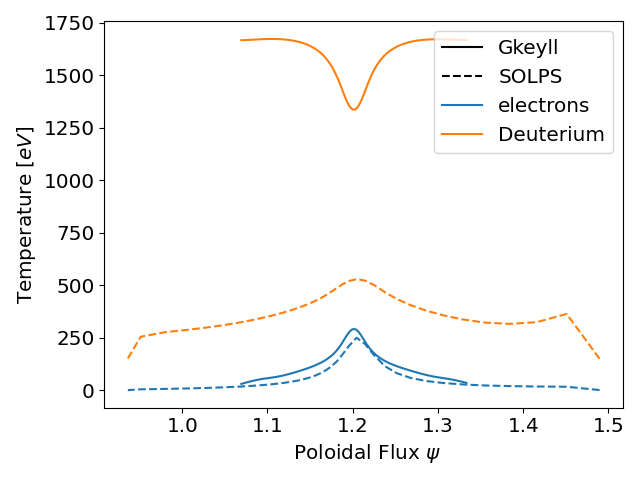}
    }
    \caption{
    Electron and ion temperature plotted along the field line at the radial center of the domain (a) and radially at the midplane (b) for Gkeyll (solid) and SOLPS (dashed) simulations in STEP geometry. The upstream ion temperature is much higher in Gkeyll because the distribution function is depleted at large $v_\parallel$ and heat leaves more slowly along the field line.
    \label{fig:temperature_plasma}
    }
\end{figure}

\begin{figure}
    \includegraphics[width=0.5\textwidth]{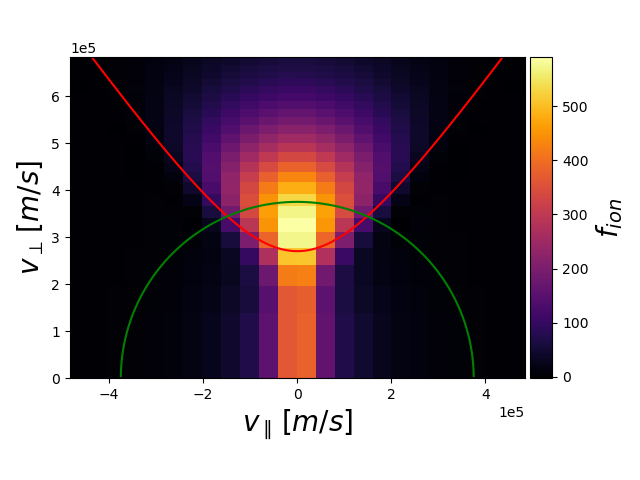}
    \caption{Ion distribution function 2mm away from the radial center at the midplane. The red line indicates the trapped-passing boundary including the effect of the potential and the green line is a contour of a Maxwellian with the same temperature as the distribution function. The ions are clearly not Maxwellian; they consist primarily of trapped particles. The depletion of the distribution function at large $v_\parallel$ is responsible for the difference in ion temperature between Gkeyll and SOLPS.
    \label{fig:ion_upstream}
    }
\end{figure}

First we compare the steady state density profiles both along the field line at the radial center and radially at the midplane shown in Fig.~\ref{fig:density_plasma}. The profiles are similar for SOLPS and Gkeyll.
However, the radial density profile is slightly wider in Gkeyll due to mirror trapping of the ions which broadens the potential and in turn broadens the electron density.
Next we examine the temperature profiles in Fig.~\ref{fig:temperature_plasma} and see an obvious difference in the ion temperatures - the Gkeyll ion temperature is much higher.
The effect responsible for this difference is mirror trapping of the ions. The mean free path is long so we should see mirror trapping of the ions upstream. The trapping will result in an ion distribution function that is depleted at large $v_\parallel$. This depletion will cause heat to move more slowly along the field line. Since heat leaves more slowly, for the same heating power, the result is a higher ion temperature in the Gkeyll simulations.
If we inspect the distribution function in Fig.~\ref{fig:ion_upstream}, we can see that the ions are clearly not Maxwellian and are instead trapped.
In Fig.~\ref{fig:ion_upstream}, the red line indicates the trapped-passing boundary
including the effect of the potential and the green line is a contour
of a Maxwellian with the same temperature as the distribution function.


An enhancement of the potential drop along the divertor leg due to the mirror force in a Super-X like divertor configuration is predicted in Ref.~\onlinecite{Mike23}. The long outer leg results in a drop in $B$ along the field line. Thus, the mirror force accelerates particles toward the divertor plate. This acceleration results in an increased average velocity parallel to the field line, $u_\parallel$, and is known as the "magnetic Laval nozzle" effect in magnetic mirrors~\cite{Xianzhu14,Andersen2003}. Particle conservation requires that $nu_\parallel/B$ remains constant, so the density must drop along the field line. If electrons have approximately a Boltzmann response, this will result in a potential drop along the field line.

\begin{figure}
    \centering
    \includegraphics[width=0.45\textwidth]{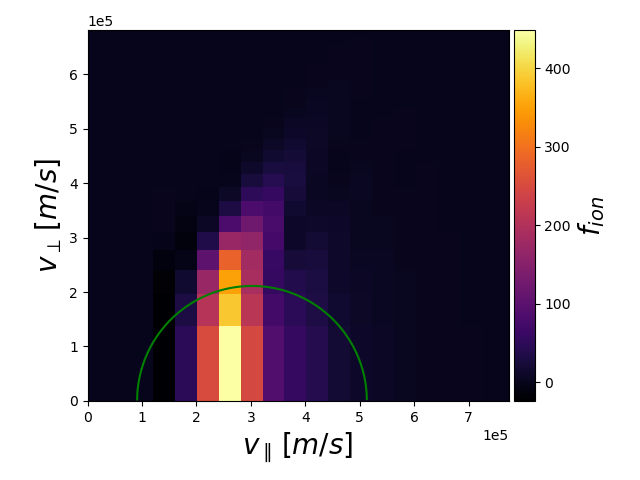}
    \caption{
    Ion distribution function downstream near the divertor plate 2mm away from the radial center. The green line is a contour of a Maxwellian with the same temperature as the distribution function. Ions with large $v_\perp$ are pushed towards higher $v_\parallel$ by the mirror force as can be seen by the way the distribution function is skewed to the right at large $v_\perp$. The mirror force is responsible for raising the ion exit velocity in Gkeyll relative to SOLPS.
    \label{fig:ion_downstream}
    }
\end{figure}

\begin{figure}
    \centering
    \includegraphics[width=0.45\textwidth]{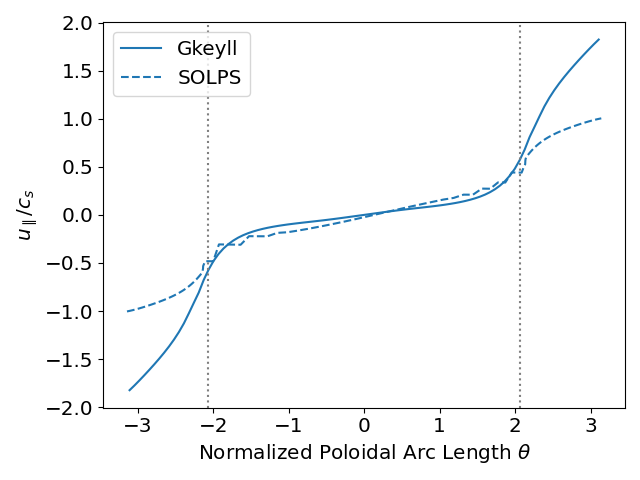}
    \caption{
    Average ion parallel velocity, $u_\parallel$, normalized to the sound speed, $c_s = \sqrt{(T_e + T_i)/m_i}$, plotted along the field line for SOLPS and Gkeyll simulations in STEP geometry. The X-points are located at $\theta  =\pm$ 2.07 (marked by the vertical gray dotted lines) and the divertor plates are located at $\theta = \pm\pi$. The ion velocity in Gkeyll exceeds that of SOLPS at the divertor plates due to acceleration by the mirror force along the divertor legs.
    \label{fig:upar_plasma}
    }
\end{figure}

\begin{figure}
    \subfloat[Potential at radial center.]{
        \includegraphics[width=0.45\textwidth]{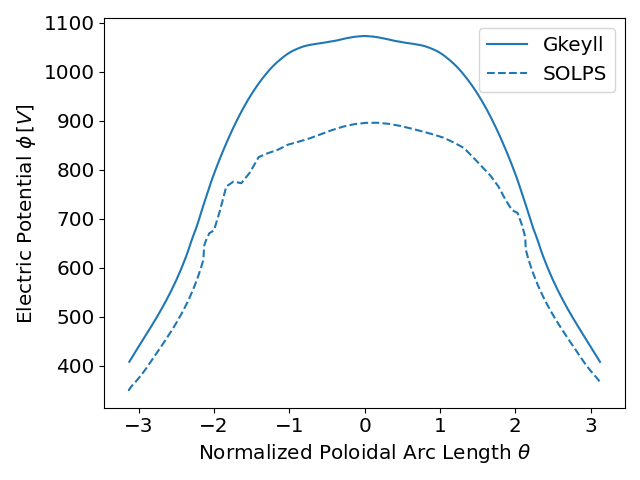}
    }\\
    \subfloat[Potential 2mm away from radial center.]{
        \includegraphics[width=0.45\textwidth]{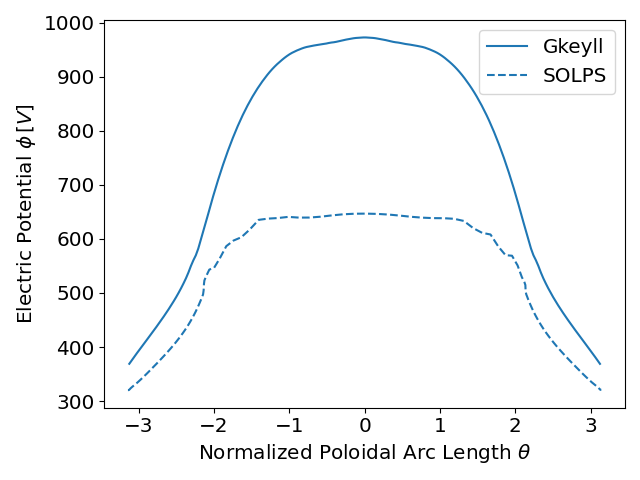}
    }
    \caption{
    Electrostatic potential, $\phi$, plotted along the field line at the radial center (a) and 2mm away (b) for SOLPS and Gkeyll simulations in STEP geometry. The potential drop from the midplane ($\theta=0$) to the divertor plate ($\theta = \pm \pi$) is 26\% larger in Gkeyll at the radial center and 80\% larger 2mm away from the radial center. The increased ion parallel velocity in Gkeyll results in this enhanced potential drop.  
    \label{fig:phi_plasma}
    }
\end{figure}

This effect can be seen clearly in figures~\ref{fig:ion_downstream},~\ref{fig:upar_plasma}, and~\ref{fig:phi_plasma}. In Fig~\ref{fig:ion_downstream} we see that high $\mu$ particles are pushed towards high $v_\parallel$. This is due to the mirror force accelerating particles towards the divertor plate. In Fig.~\ref{fig:upar_plasma} we see that in the region beyond the X-point, the Gkeyll $u_\parallel$ exceeds that of SOLPS as predicted. In SOLPS, the Bohm condition, $u_\parallel = c_s$ where $c_s = \sqrt{(T_e + T_i)/m_i}$ is the local sound speed, is enforced at the divertor plate, but in Gkeyll the ion outflow is supersonic.
This effect has also been explored in simplified magnetic geometries in Refs.~\onlinecite{Xianzhu14} and~\onlinecite{Sabo2022}.
Finally, we see the effect of the increased ion outflow velocity on the potential in Fig.~\ref{fig:phi_plasma}. At the radial center, the effect is moderate with a 26\% larger potential drop from midplane to divertor plate in Gkeyll. However, 2mm away from the radial center where the collisionality is lower, the effect is very large; the Gkeyll potential drop is 80\% larger.


\section{Simulations with Argon Impurities}\label{sec:impurities}

\subsection{10 eV Neutral Argon up to Ar$^{4+}$}
\label{sec:ar4}
In order to see how the low impurity temperature along with the increased potential drop produced by the mirror force affects impurity confinement, we added a static 10 eV neutral argon background localized near the divertor plate. The neutral argon profile is shown in Fig.~\ref{fig:Ar0}.
In the initial conditions of the simulation, the density of all charged argon states is zero. The only sources of Ar$^{+1}$ are ionization of neutral argon and recombination of Ar$^{+2}$, the only sources of Ar$^{2+}$ are ionization of Ar$^{+1}$ and recombination of Ar$^{+3}$, and so on.

The neutral argon profiles in SOLPS and Gkeyll are identical, so differences in the steady state profiles for argon charge states 1 through 4 can be attributed to the difference in the potential drop described in the previous section as well as the difference in argon and ion temperatures included in Gkeyll. In SOLPS, the argon is assumed to be the same temperature as the main ion species, deuterium. This assumption is not a choice made by the user; SOLPS has only one ion energy equation, so there is only one ion temperature.

In Ref.~\onlinecite{Mike23}, it is argued that the assumption of equal main ion and impurity temperature will not be valid if the mean free path is long. The heating rate for a cold impurity species by deuterium is given by 
\begin{equation}
\frac{d T_Z}{d t} \sim v_D \frac{m_D}{m_Z} Z^2 T_D,
\end{equation}
where $\nu_D$ is the deuterium-deuterium collision frequency.
A simple estimate for the time, $\tau$, for an impurity to be expelled by the potential $\phi$ over a length $L_\parallel$ is
\begin{equation}
\tau \sim L_{\|}\left(Z e \phi / m_Z\right)^{-1 / 2}
\end{equation}

Heating over this time will give an impurity temperature $T_Z \sim \tau \dv{T_z}{t}$ giving a temperature ratio of
\begin{equation}
\frac{T_Z}{T_D} \sim L_{\|} Z^2 v_D \frac{m_D}{m_Z}\left(Z e \phi / m_Z\right)^{-1 / 2}
\label{eq:heating_1}
\end{equation}

From the simulation results, we see that $e\phi\sim T_D$. Substituting this and $\nu_D = \sqrt{T_D/m_D}/\lambda_{mfp,D}$ into Eq.~\ref{eq:heating_1}, we get
\begin{equation}
\frac{T_Z}{T_D} \sim\left(\frac{L_{\|}}{\lambda_{m f p, D}}\right)\left(\frac{m_D}{m_Z}\right)^{1 / 2} Z^{3 / 2}
\label{eq:temp_ratio}
\end{equation}

In the regime we are simulating, the first term, $L_\parallel/\lambda_{mfp,D}$ is small; it is approximately 0.02 near the divertor plate at the radial center where the impurities are concentrated. The mass ratio will be small for impurities with a large atomic number like argon. The heating rate also scales with the charge as $Z^{3/2}$, so higher charge state impurities will be heated more quickly.

If impurities are much colder than the deuterium and $e\phi \sim T_D$, the amount of impurities able to overcome the potential barrier and make it upstream will be small. This shielding effect will be stronger for larger mass and lower charge impurities.

\begin{figure}
    \includegraphics[width=\columnwidth]{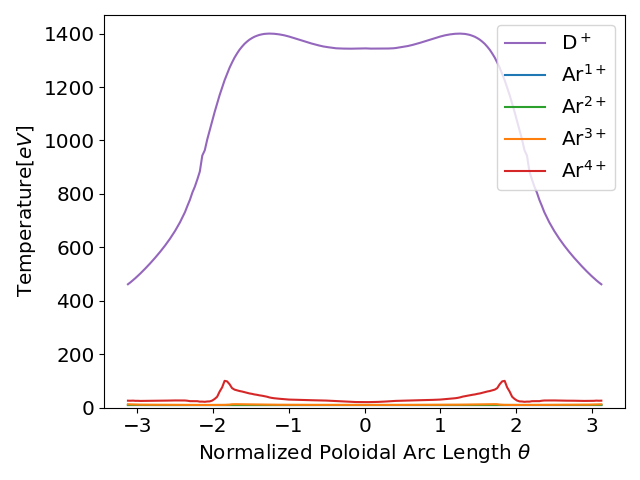}
    \caption{
    Deuterium and argon charge state temperatures plotted along the field line at the radial center for the Gkeyll simulation with 10eV neutral argon and charge states up to Ar$^{4+}$. All of the argon charge states are much colder than the deuterium, which invalidates the assumption of equal ion and impurity temperature made by SOLPS. The argon is expelled by the potential before it has time to thermally equilibrate with the deuterium.
    \label{fig:temperature_ar}
    }
\end{figure}

In Fig.~\ref{fig:temperature_ar}, we can see that the assumption of equal deuterium and impurity temperatures is invalidated by Gkeyll; the assumption is inaccurate this kinetic regime. Only the Ar$^{4+}$ is significantly heated, but even the Ar$^{4+}$ temperature is drastically lower than the deuterium temperature. We also see that the downstream Ar$^{4+}$ temperature is lower than the upstream temperature; the downstream temperature is what will determine how much Ar$^{4+}$ can make it upstream.

\begin{figure}
    \includegraphics[width=\columnwidth]{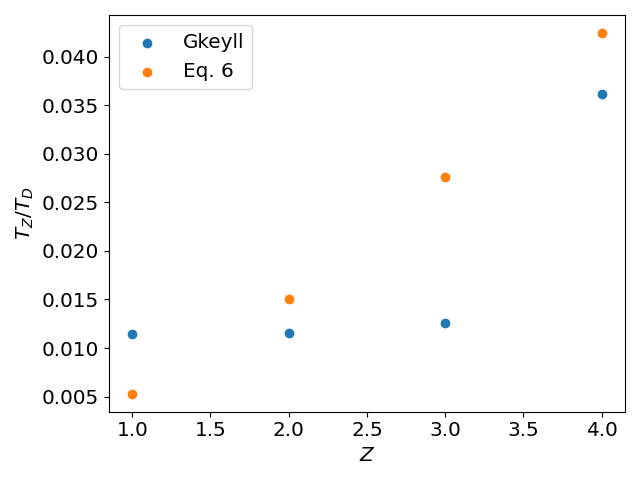}
    \caption{
        Poloidally averaged ratio of argon charge state temperature to deuterium temperature predicted by Eq.~\ref{eq:temp_ratio} and observed in the simulation plotted at the radial center of the simulation domain vs. charge state (Z) for the Gkeyll simulation with 10eV neutral argon and charge states up to Ar$^{4+}$. The predicted ratio agrees quite well with the observed ratio for the highest charge state, Ar$^{4+}$ but, as expected, overestimates the ratio for several of the lower charge states. Lower charge states are ionized quickly and do not have as much time to be heated as Eq.~\ref{eq:temp_ratio} assumes.
    \label{fig:temp_ratio}
    }
\end{figure}

In Fig.~\ref{fig:temp_ratio}, we compare the argon charge state temperatures observed in Gkeyll with those predicted by Eq.~\ref{eq:temp_ratio}. It is important to note that Eq.~\ref{eq:temp_ratio} takes into account losses due to expulsion by the potential but not due to ionization. 
Lower charge states are ionized very quickly at the electron temperatures observed in these simulations, so they have less time to be heated than assumed by Eq.~\ref{eq:temp_ratio}. Thus, Eq.~\ref{eq:temp_ratio} will overestimate the temperature of low charge states. In Fig.~\ref{fig:temp_ratio} we plot the ratio $T_Z/T_D$ for each argon charge state at the radial center of the simulation domain averaged in the parallel direction along with the average temperature predicted by Eq~\ref{eq:temp_ratio}. The predicted ratio agrees quite well with the observed ratio for the highest charge state, Ar$^{4+}$ but, as expected, overestimates the ratio for several of the lower charge states. In the next section, we will compare the predicted and observed temperature ratio for more charge states and the trend will become clearer.

\begin{figure}
    \centering
    \subfloat[\label{fig:density_ar_total}Neutral and charged argon density (summed over charge states 1 through 4) plotted along the field line in Gkeyll and SOLPS.]{
        \includegraphics[width=0.45\textwidth]{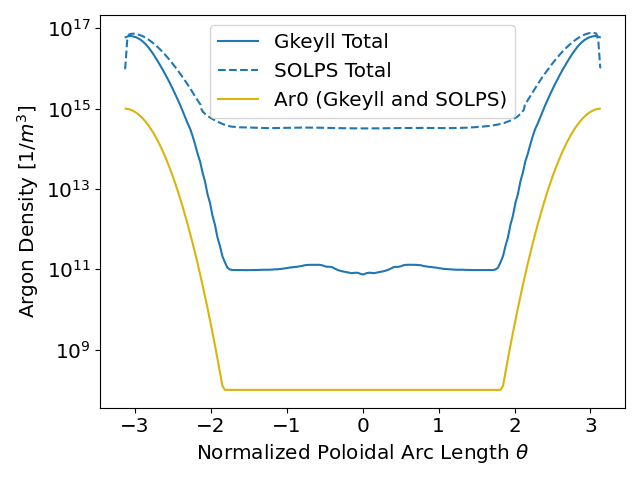}
    }\\
    \subfloat[\label{fig:density_ar_individual}Neutral, Ar$^{+1}$, and Ar$^{+4}$ density plotted along the field line in Gkeyll and SOLPS.]{
        \includegraphics[width=0.45\textwidth]{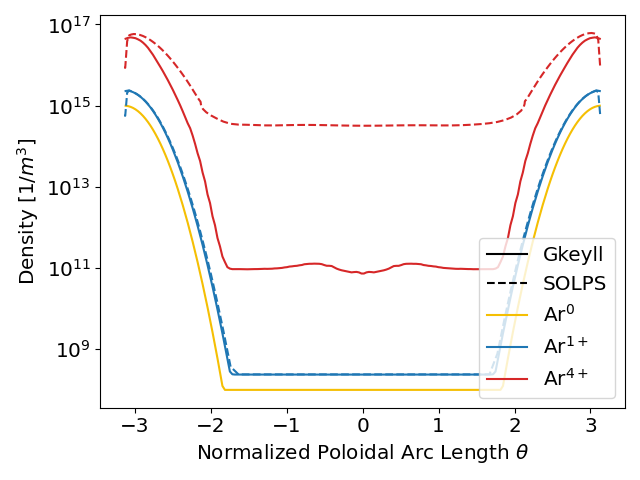}
    }
    \caption{Argon density plotted along the field line at the radial center in SOLPS and Gkeyll for the Gkeyll simulation with 10 eV neutral argon and charge states up to Ar$^{4+}$. The downstream density of charged argon is similar, but Gkeyll's upstream density is orders of magnitude lower.}
\end{figure}

Next we can see the effect of this temperature difference as well as the enhanced potential drop on the argon density profiles. In Fig.~\ref{fig:density_ar_total}, we plot the neutral argon density along with the total charged argon density, summed over charge states 1 through 4, along the field line for both SOLPS and Gkeyll. The shielding effect is quite clear; the total charged argon density in Gkeyll is much lower upstream. The charged argon density is similar downstream near the divertor plate, but Gkeyll's upstream argon density is four orders of manitude lower. This drastic difference can be attributed to the shielding effect created by the enhanced potential drop and the fact that the argon is much colder than the deuterium.

In Fig.~\ref{fig:density_ar_individual}, we look at some of the individual charge state densities for both Gkeyll and SOLPS. We can see here that Ar$^{4+}$ dominates the density both upstream and downstream. In light of Fig.~\ref{fig:temperature_ar} and Eq.~\ref{eq:temp_ratio}, it makes sense that higher charge states should dominate upstream.
Only high charge states are heated enough overcome the potential barrier and travel upstream. Furthermore, at these electron temperatures, any low charge states that do make it upstream will be quickly ionized. The Ar$^{+1}$ profiles in Gkeyll and SOLPS agree quite well both downstream and upstream because, at these electron temperatures, the steady state Ar$^{+1}$ density is set almost entirely by the ionization rate of neutral argon and Ar$^{+1}$.


\subsection{10 eV Neutral Argon up to Ar$^{8+}$}
\label{sec:ar8}
One may think that higher charge states will acquire a significant density upstream because they will be hotter and more capable of overcoming the potential barrier, but this does not happen. 
The reason is that higher charge states must come from ionization of lower charge states, which are strongly shielded and confined to the divertor region.

High Z impurities are expelled very strongly by the potential since the electrostatic force is proportional to Z. Low Z impurities are strongly shielded since they do not have time to heat up and will not make it upstream to be ionized to a higher charge state. Higher charge states originate from ionization of lower chage states, so the shielding of low charge states should result in a lower concentration of high charge states even if the higher charge states are thermally equilibrated. In this section we provide evidence to support this argument; we ran simulations identical to those described in the previous subsection~\ref{sec:ar4} but included up to  Ar$^{8+}$.

\begin{figure}
    \centering
    \includegraphics[width=\columnwidth]{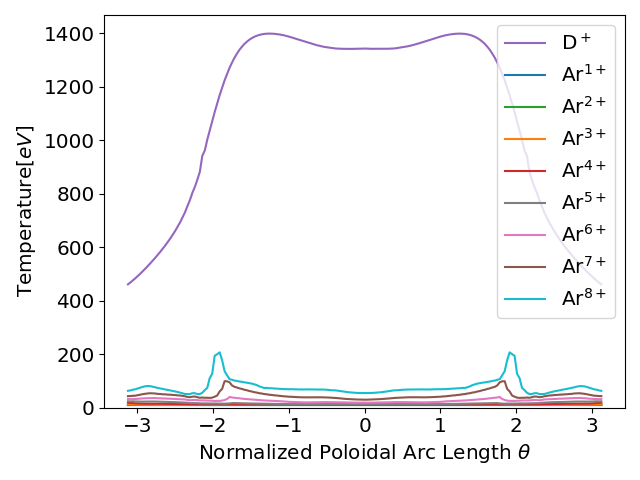}
    \caption{Deuterium and argon charge state temperatures plotted along the field line at the radial center for the Gkeyll simulation with 10eV neutral argon and charge states up to Ar$^{8+}$. Higher argon charge states have higher temperatures, but their temperature is still much lower than the deuterium temperature.}
    \label{fig:temperature_ar8}
\end{figure}
\begin{figure}
    \includegraphics[width=\columnwidth]{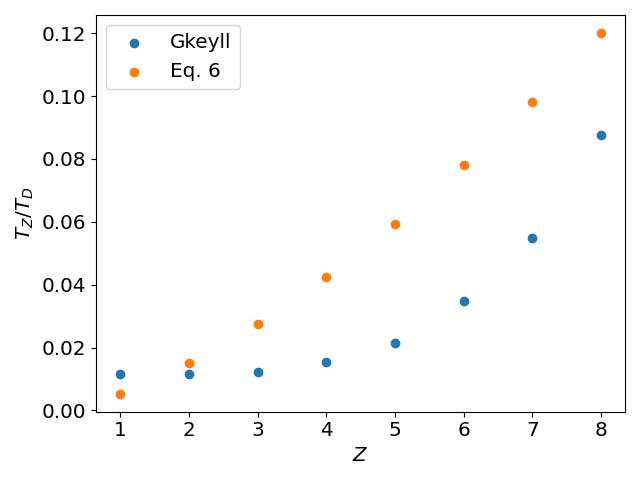}
    \caption{
        Poloidally averaged ratio of argon charge state temperature to deuterium temperature predicted by Eq.~\ref{eq:temp_ratio} and observed in the simulation plotted at the radial center of the simulation domain vs. charge state (Z) for the Gkeyll simulation with 10eV neutral argon and charge states up to Ar$^{8+}$. The predicted ratio agrees quite well with the observed ratio for the highest charge state, Ar$^{8+}$ but, as expected, overestimates the ratio for most of the lower charge states. Lower charge states are ionized quickly and do not have as much time to be heated as Eq.~\ref{eq:temp_ratio} assumes.
    \label{fig:temp_ratio_ar8}
    }
\end{figure}

In Fig.~\ref{fig:temperature_ar8}, we plot the argon charge state temperatures along the field line. Ar$^{4+}$ is not as hot as it was in the simulation which included only up to charge state 4 because ionization of Ar$^{4+}$ reduces the amount of time available for it to be heated. Higher charge states are indeed hotter than lower charge states as predicted by Eq.~\ref{eq:temp_ratio}. As observed in the previous section, the downstream temperature is significantly lower than the upstream temperature.
In Fig.~\ref{fig:temp_ratio_ar8} we plot the ratio $T_Z/T_D$ for each argon charge state at the radial center of the simulation domain averaged in the parallel direction along with the average temperature predicted by Eq~\ref{eq:temp_ratio}. The predicted ratio agrees quite well with the observed ratio for the highest charge state, Ar$^{8+}$ but again overestimates the ratio for several of the lower charge states.

\begin{figure}
    \subfloat[\label{fig:density_ar8_total}Neutral and charged argon density (summed over charge states 1 through 8) plotted along the field line in Gkeyll and SOLPS.]{
        \includegraphics[width=0.45\textwidth]{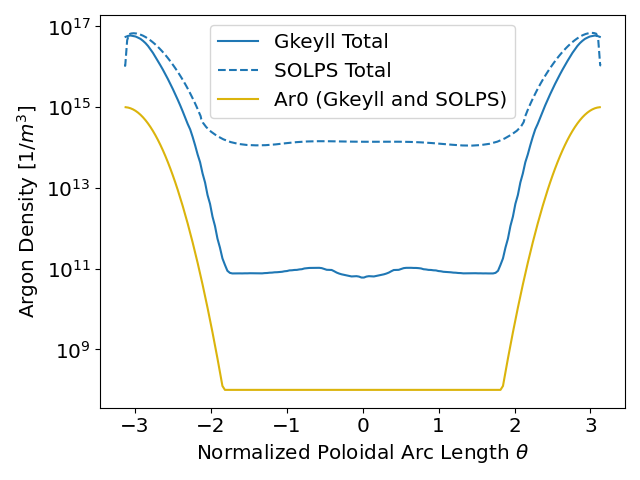}
    }\\
    \subfloat[\label{fig:density_ar8_individual} Neutral, Ar$^{+1}$, Ar$^{+5}$, and Ar$^{+8}$ density plotted along the field line in Gkeyll and SOLPS.]{
        \includegraphics[width=0.45\textwidth]{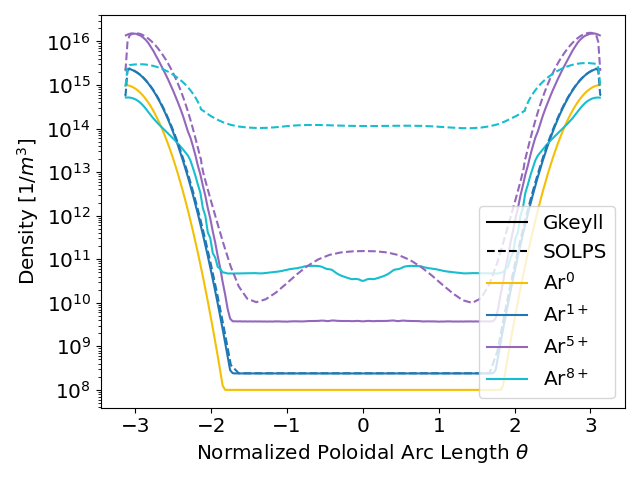}
    }
    \caption{Argon density plotted along the field line in SOLPS and Gkeyll for the Gkeyll simulation with 10eV neutral argon and charge states up to Ar$^{8+}$. Charge state 5 dominates near the divertor plate because charge states higher than 5 are pushed out more quickly by  the potential. The total argon density is nearly identical to that of the simulation which included only up to Ar$^{4+}$.
    }
\end{figure}

In Fig.~\ref{fig:density_ar8_individual} we plot some of the individual charge state densities. In both codes, Ar$^{5+}$ dominates near the divertor plate and higher charge states such as Ar$^{8+}$ have a lower downstream density. This can be attributed to the fact that higher charge states are expelled more quickly by the potential. Both codes also show that Ar$^{8+}$ dominates upstream which also makes sense; in Gkeyll only higher charge states should be able to travel upstream and in both codes any low charge states that make it upstream will be quickly ionized.

In Fig.\ref{fig:density_ar8_total}, we plot the total argon density summed over charge states 1 through 8 along the field line for both SOLPS and Gkeyll. This figure supports the argument that the strong shielding effect for low charge states does in fact reduce the upstream density of high charge states. Comparing Fig.\ref{fig:density_ar8_total} and Fig.~\ref{fig:density_ar_total} we see that the upstream and downstream total charged Argon density is nearly identical when we include up to Ar$^{4+}$ or Ar$^{8+}$. So, the inclusion of higher charge states, which heat faster, did not have significant effect on the total upstream or downstream impurity density.

In Gkeyll, including charge states beyond Ar$^{8+}$ is computationally expensive. However, in SOLPS, we ran another simulation including up to Ar$^{12+}$, not shown here, which confirmed that the inclusion of higher charge states did not have a significant effect on the total upstream or downstream argon density.

\subsection{500 eV Neutral Argon}
\label{sec:hotar}
In  order to verify that the difference in upstream argon densities is primarily caused by the difference in argon temperature, 
we ran another Gkeyll simulation identical to the one described in the section~\ref{sec:ar4} except for an increase in the neutral argon temperature to 500 eV. We chose this temperature because the deuterium ion temperature (and thus the argon ion temperature) in SOLPS is approximately 500 eV. The argon density in this simulation is sufficiently low that the change in argon temperature from 10 to 500 eV does not significantly affect the deuterium or electron profiles. In this case, we expect differences in the argon density between SOLPS and Gkeyll to be drastically reduced. 

\begin{figure}
    \includegraphics[width=\columnwidth]{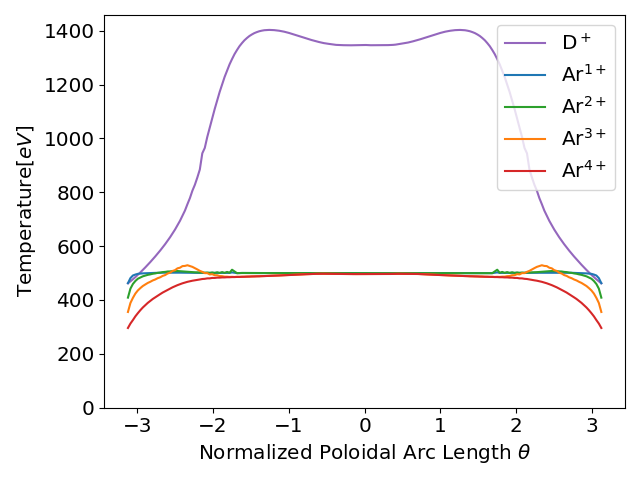}
    \caption{
    Deuterium and argon temperature plotted along the field line in Gkeyll for the simulation with 500 eV neutral argon and charge states up to Ar$^{4+}$. In this case, the argon temperature near the target is similar to the deuterium temperature. This is of course expected based on our choice of the neutral argon temperature.
    \label{fig:temperature_arhot}
    }
\end{figure}

In Gkeyll we cannot set the temperature of the charged argon species equal to the deuterium temperature or hold it static, so there is still a difference between the argon and deuterium temperatures as seen in Fig.~\ref{fig:temperature_arhot}. This could be responsible for some differences in the argon density between the two codes,  and the difference in potential drop will also contribute.

\begin{figure}
    \includegraphics[width=\columnwidth]{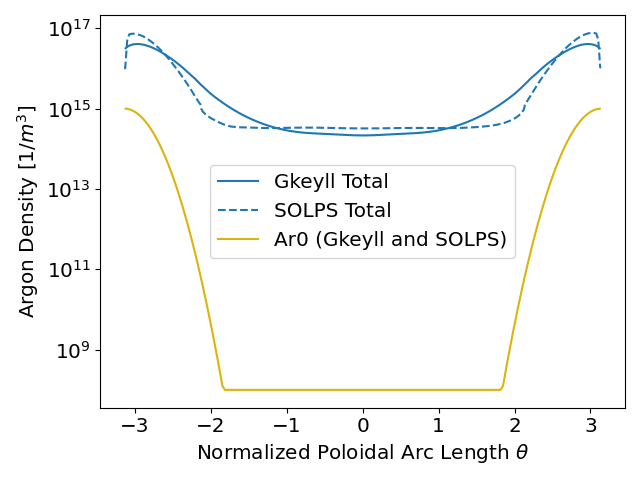}
    \caption{
    Neutral and charged argon density (summed over charge states 1 through 4) plotted along the field line in SOLPS and Gkeyll for the Gkeyll simulation with 500 eV neutral argon and charge states up to Ar$^{4+}$. In this case the upstream argon densities in Gkeyll and SOLPS are only a factor of 2 different. This is expected and indicates that the low impurity temperature is primarily responsible for the huge differences in upstream density we saw with 10eV neutral argon.
    \label{fig:density_arhot_total}
    }
\end{figure}

As seen in Fig.~\ref{fig:density_arhot_total},
the argon densities agree much better when the neutral argon temperature is set to 500 eV in Gkeyll. 
The downstream total argon density is similar to SOLPS as it was in the 10 eV case. The difference between the 10 eV and 500 eV case can be seen in the upstream argon density. In the 10 eV case Gkeyll's upstream density was a factor of $10^4$ lower than SOLPS's but in the 500 eV case Gkeyll's upstream density is only a factor of 2 lower. This indicates that the low impurity temperature is the primary factor responsible for the huge differences in upstream argon density seen in the previous sections.

\subsection{Higher Radiation Fractions}
\label{sec:high_frac}

The simulations we have shown so far have low enough argon density that radiation is negligible ($<1\textrm{ MW}$) and the deuterium and electron profiles are unaffected. In a practical scenario, most of the electron power will need to be radiated. In order to verify that the superior impurity confinement seen in kinetic simulations holds in regimes with higher radiation fraction, we ran one final set of simulations.

\begin{figure}
    \subfloat[ \label{fig:density_arhigh_total} Charged argon density (summed over charge states 1 through 4) plotted along the field line at the radial center in Gkeyll and SOLPS.
    ]{
        \includegraphics[width=0.45\textwidth]{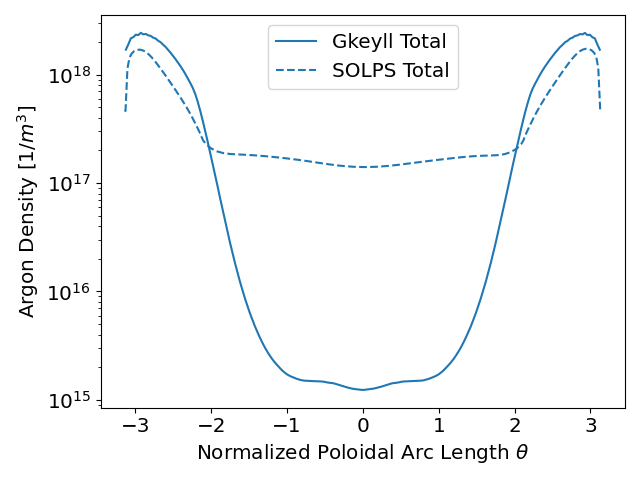}
    }\\
    \subfloat[\label{fig:high_frac_artemp}
    Deuterium and argon temperature plotted along the field line at the radial center in Gkeyll.]{
        \includegraphics[width=0.45\textwidth]{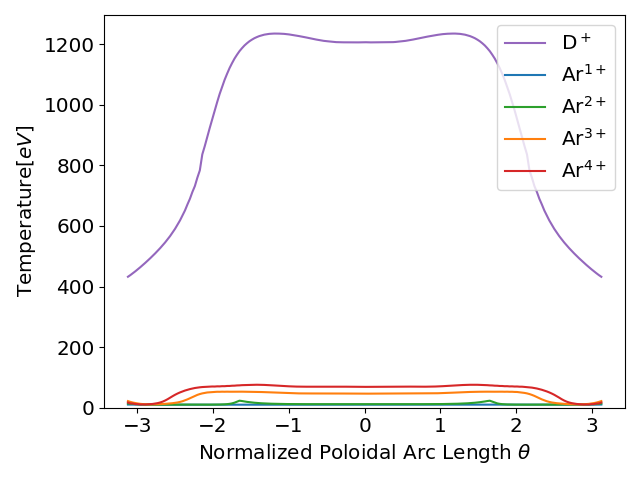}
    }
    \caption{
    Argon densities (a) and deuterium and argon temperatures (b) from the high radiation simulations with argon charge states up to Ar$^{4+}$. The neutral argon density at the divertor plate is 1.5$\times$10$^{16}$ in Gkeyll and  1$\times$10$^{16}$ in SOLPS. This results in 31MW of radiation in Gkeyll and 27 MW in SOLPS. The upstream argon density in Gkeyll is a factor of 7 lower.
    }
\end{figure}

This set of simulations is similar to the set described in subsection~\ref{sec:ar4}, but instead of matching the neutral argon densities in SOLPS and Gkeyll, we scaled the neutral argon profiles until we achieved a similar amount of total radiated power in the two codes. 
We used a peak neutral argon density of 1$\times$ 10$^{16}$ in SOLPS and 8.1$\times$ 10$^{15}$ in Gkeyll. The neutral argon temperature is 10 eV. This resulted in 27 MW of radiated power in SOLPS and 35 MW in Gkeyll. The total upstream argon density in Gkeyll for this case is two orders of magnitude lower than in SOLPS as seen in Fig.~\ref{fig:density_arhigh_total}. In total, 40MW of power are put into the electrons, so 25MW of radiation constitutes a significant radiation fraction.

In this high radiation case we see that, while still very large, the difference in upstream impurity density between SOLPS and Gkeyll is reduced relative to the low radiation case; the difference was a factor of 10$^4$ in the low radiation case and is a factor of  10$^2$ here.
However, this case shows that the shielding effect is still very strong even at higher radiation fractions; there is still drastically superior confinement of impurities to the divertor region in Gkeyll.

\begin{figure}
    \subfloat[\label{fig:high_frac_temp} Electron and deuterium temperatures from SOLPS and Gkeyll plotted along the field line at the radial center.]{
    \includegraphics[width=0.4\textwidth]{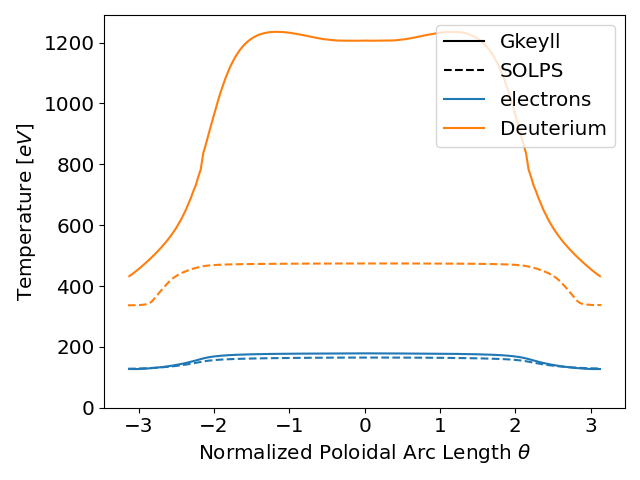}
    }\\
    \subfloat[ \label{fig:high_frac_phi} Electrostatic potential from SOLPS and Gkeyll plotted along the field line at the radial center. ]{
    \includegraphics[width=0.4\textwidth]{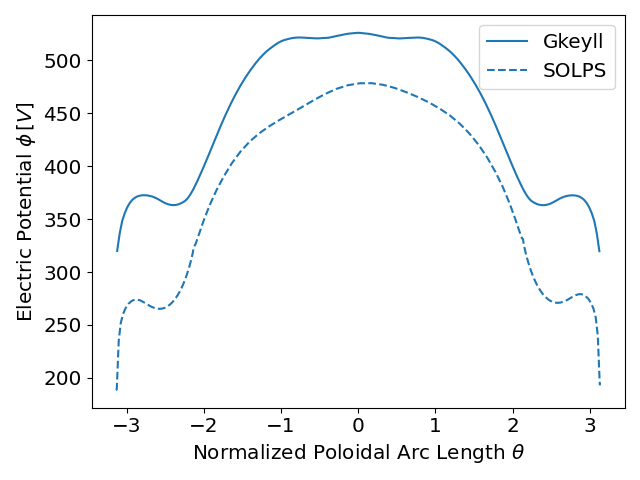}
    }\\
    \subfloat[ \label{fig:high_frac_density} Electron density from SOLPS and Gkeyll plotted along the field line at the radial center. ]{
    \includegraphics[width=0.4\textwidth]{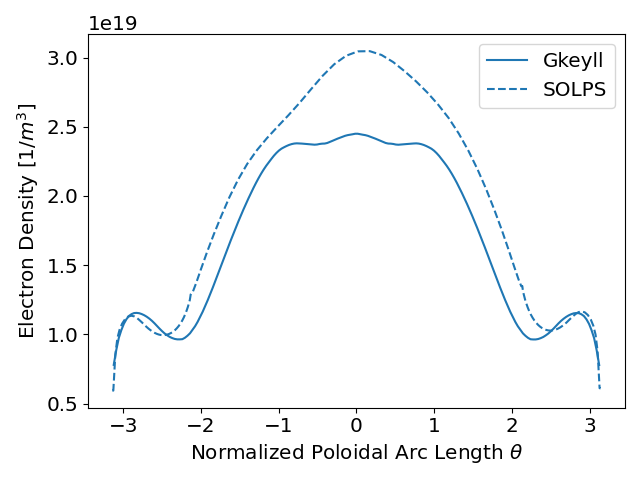}
    }
    \caption{
    Electron and deuterium temperatures (a), electrostatic potential (b), and electron density (c) from the high radiation simulations with argon charge states up to Ar$^{4+}$. The electron temperature is greatly reduced near the divertor plate relative to the simulations without impurities. The reduction in electron temperature results in a reduction in the potential drop from midplane to divertor plate relative to simulations without impurities. The upstream electron  density in Gkeyll is lower than in SOLPS as a result of the lower upstream impurity density.
    }
\end{figure}

We also see that the electron temperature is significantly reduced by the radiation; it is reduced by a factor of 2 downstream relative to the simulation without impurities as can be seen by comparing Fig.~\ref{fig:high_frac_temp} to Fig.~\ref{fig:temperature_plasma_z}. The electrostatic potential drop from midplane to divertor plate is proportional to the electron temperature, so the drop in electron temperature results in a reduction in the potential drop as can be seen by Fig.~\ref{fig:high_frac_phi} to Fig.~\ref{fig:phi_plasma} (a).

In this simulation, the upstream electron density in Gkeyll is significantly lower than in SOLPS as shown in Fig.~\ref{fig:high_frac_density}. Lower upstream impurity densities help maintain lower upstream densities for the main plasma which can be beneficial for confinement~\cite{Mike23}.
These results in this section indicate that kinetic effects may enable more effective strategies for reactor scenarios to radiate power while avoiding core contamination and maintaining good confinement.



The high radiation Gkeyll simulation is not completely converged as we did not have enough time to run it to convergence. However, most quantities from the simulation have converged, and we believe the current results are resonable. Time traces of various quantities from this simulation are included in Appendix~\ref{sec:convergence}.

\section{Conclusion}\label{sec:conclusion}
We have conducted a comparison of axisymmetric fluid and gyrokinetic simulations of a proposed STEP SOL using SOLPS and Gkeyll. We first established baseline agreement between the codes in a slab geometry in section~\ref{sec:slab_comparison}. We then demonstrated in section~\ref{sec:plasma_only} that, with upstream parameters typical for a pilot plant, kinetic effects play an important role. Mirror trapping of the ions results in a much higher ion temperature in kinetic simulations. We also found that a super-X like divertor with a long outer leg results in an enhanced ion outflow speed and electrostatic potential drop which can be seen in Gkeyll but not SOLPS because SOLPS excludes the mirror force.

In section~\ref{sec:impurities} we included argon impurities in both codes and found that the assumption of equal main ion and impurity temperature made in SOLPS is violated. In agreement with predictions made in Ref.~\onlinecite{Mike23}, the low impurity temperature prevents impurities from travelling upstream in kinetic simulations. With identical neutral density profiles at low radiation fractions, kinetic simulations achieved a similar downstream impurity density to fluid simulations but maintained an upstream impurity density which was orders of magnitude lower. At high radiation fractions, kinetic simulations still demonstrated much lower upstream impurity densities.


These findings have important implications for reactor relevant regimes. 
Higher SOL ion temperatures observed in kinetic simulations might lead to higher pedestal ion temperature which can result in higher core fusion power. Higher ion temperatures can also increase wall erosion by increasing the energy of neutrals hitting the main chamber wall, due to charge exchange of the hot plasma with cold recycled neutrals.
Our results also indicate that we may be able to support much larger downstream impurity densities (and hence more radiated power) while avoiding unacceptable upstream densities than previously thought based on SOLPS simulations. Better confinement of impurities to the divertor region entails at least two benefits: (1) avoidance of impurity contamination of the core plasma, and (2) avoidance of high upstream densities, which can degrade confinement.

We would like to mention that, while much more expensive than a fluid model, Gkeyll's axisymmetric kinetic model is not prohibitively expensive; it is inexpensive enough that it can be used for divertor studies and to conduct parameter scans. Most of the simulations conducted in this work took 1-3 days on a modest number of GPUs. Details of the runtime and cost for each simulation can be found in appendix~\ref{sec:cost}.

In future work, we will explore more realistic scenarios by including dynamic neutrals, recycling and charge exchange in the scenario described in section~\ref{sec:high_frac}. Charge exchange will be essential for removing power from the ions and reducing the heat load to the divertor plates to acceptable levels. We will also explore the effects of drifts in axisymmetric simulations.

\begin{acknowledgments}
This work was supported by the U.S. Department of Energy through the SciDAC collaboration Computational Evaluation and Design of Actuators for Core-Edge Integration (CEDA) under contract number DE-AC02-09CH11466 and DE-FG02-04ER54742. The United States Government retains a non-exclusive, paid-up, irrevocable, world-wide license to publish or reproduce the published form of this manuscript, or allow others to do so, for United States Government purposes. This work has been partly funded by STEP, a major technology and infrastructure programme led by UK Industrial Fusion Solutions Ltd (UKIFS), which aims to deliver the UK's prototype fusion powerplant  and a path to the  commercial viability of fusion.
\end{acknowledgments}

\appendix
\section{Simulation Details}
\label{sec:resolution}
In this section we detail the resolution of each Gkeyll simulation we conducted. The simulation with slab geometry described in section~\ref{sec:slab_comparison} used 72 cells in the radial direction and 64 cells in the parallel direction ($N_R\times N_Z = 72x64$). All simulations in the ST geometry had the same resolution in configuration space: 40 cells in the radial direction ($N_\psi = 40$) and 96 cells in the parallel direction ($N_\theta = 96$). The details of the velocity space resolution can be found in tables~\ref{tab:resolution} and~\ref{tab:extents}.

The version of the Gkeyll code used was commit 5d78312 on the gk-g0-app branch of \url{https://github.com/ammarhakim/gkylzero}. In order to make the simulation results reproducible, we have made the input files, which contain all of the details of the simulation setup, for the simulations conducted in this work available at \url{https://github.com/ammarhakim/gkyl-paper-inp/tree/master/2024_PoP_GKST}. 

\begin{table*}
\caption{\label{tab:resolution} Velocity space resolution for each species in each simulation. Each entry in the table gives the velocity space resolution ($N_{v_\parallel} \times N_\mu$) for the species in the corresponding column in the simulation in the corresponding row.}
\begin{ruledtabular}
\begin{tabular}{ccccccccccc}
 Section&electrons&$D^+$&$Ar^{+1}$&$Ar^{+2}$&$Ar^{+3}$&$Ar^{+4}$&$Ar^{+5}$&$Ar^{+6}$&$Ar^{+7}$&$Ar^{+8}$\\ \hline
 ~\ref{sec:slab_comparison}&16x12&16x12&--&--&--&--&--&--&--&--\\
 ~\ref{sec:plasma_only}&16x12&16x12&--&--&--&--&--&--&--&--\\
 ~\ref{sec:ar4}&16x12&16x12&16x12&16x12&16x12&16x12&--&--&--&--\\
 ~\ref{sec:ar8}&16x12&16x12&16x12&16x12&16x12&16x12&16x12&16x12&16x12&16x12\\
 ~\ref{sec:hotar}&16x12&16x12&16x12&16x12&16x12&16x12&--&--&--&--\\
 ~\ref{sec:high_frac}&16x12&16x12&16x12&32x12&48x12&96x12&--&--&--&--\\
\end{tabular}
\end{ruledtabular}
\end{table*}

\begin{table*}
\caption{\label{tab:extents} Parallel velocity extents for each species in each simulation. Each entry lists the reference temperature, $T_{ref}$, in eV, the maximum parallel velocity, $v_{\parallel,max}$, normalized to $v_t = \sqrt{T_{ref}/m}$, and the maximum magnetic moment, $\mu_{max}$, normalized to $mv_t^2/2B_0$. $B_0$=2.51 for all simulations.}
\begin{ruledtabular}
\begin{tabular}{ccccccccccc}
 Section&electrons&$D^+$&$Ar^{+1}$&$Ar^{+2}$&$Ar^{+3}$&$Ar^{+4}$&$Ar^{+5}$&$Ar^{+6}$&$Ar^{+7}$&$Ar^{+8}$\\ \hline
 ~\ref{sec:slab_comparison}&63,6,12&94,6,12&--&--&--&--&--&--&--&--\\
 ~\ref{sec:plasma_only}&364, 6, 18&546, 6, 18&--&--&--&--&--&--&--&--\\
 ~\ref{sec:ar4}&364, 6, 18&546, 6, 18&10, 4, 18&10, 8, 18&10, 12, 36&10, 22, 72&--&--&--&--\\
 ~\ref{sec:ar8}&364, 6, 18&546, 6, 18&10, 4, 18&10, 8, 18&10, 12, 36&10, 22, 72&10, 6, 72&110,8.5,144&10, 9, 144&10,9,144\\
 ~\ref{sec:hotar}&364, 6, 18&546, 6, 18&500, 6, 18&500, 6, 18&500, 6, 18&500, 6, 18&--&--&--&--\\
 ~\ref{sec:high_frac}&364, 6, 18&546, 6, 18&10, 4, 18 &10, 8, 18&10, 12, 18&10, 24, 18&--&--&--&--\\
\end{tabular}
\end{ruledtabular}
\end{table*}

\section{Simulation Cost}
\label{sec:cost}
The Gkeyll simulation section~\ref{sec:slab_comparison} was run on 2 nodes (4 GPUs per node) and all other Gkeyll simulations were run on 8 nodes. Table~\ref{tab:runtime} lists the number of node hours used and the final simulation time for each Gkeyll simulation. Simulations with impurities are started with initial conditions for the deuterium and electrons given by the steady of the plasma only simulation in section~\ref{sec:plasma_only}.  

\begin{table}
\caption{\label{tab:runtime} Simulation cost and end time.}
\begin{ruledtabular}
\begin{tabular}{ccccccccccc}
 Section&Node Hours& End time  (ms)\\ \hline
 ~\ref{sec:slab_comparison}&96&4\\
 ~\ref{sec:plasma_only}&160&8\\
 ~\ref{sec:ar4}&256&1.83\\
 ~\ref{sec:ar8}&576&2.16\\
 ~\ref{sec:hotar}&384&2.75\\
 ~\ref{sec:high_frac}&3648&25.7\\
\end{tabular}
\end{ruledtabular}
\end{table}

\section{Convergence of  the High Radiation Case}
\label{sec:convergence}
\begin{figure}
    \subfloat[\label{fig:trace_density} ]{
    \includegraphics[width=0.4\textwidth]{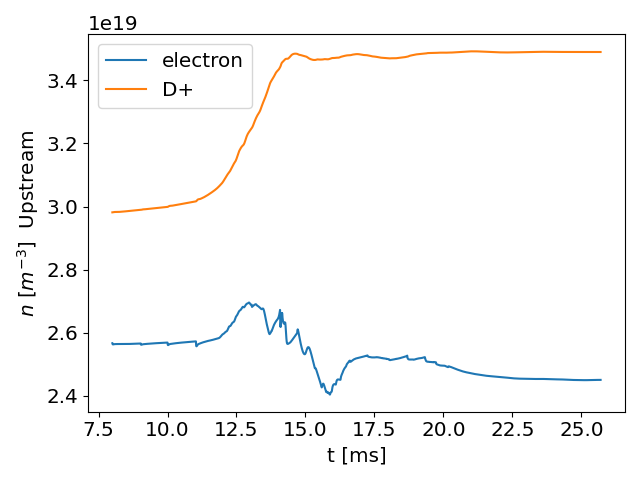}
    }\\
    \vspace{-1cm}
    \subfloat[\label{fig:trace_density_down} ]{
    \includegraphics[width=0.4\textwidth]{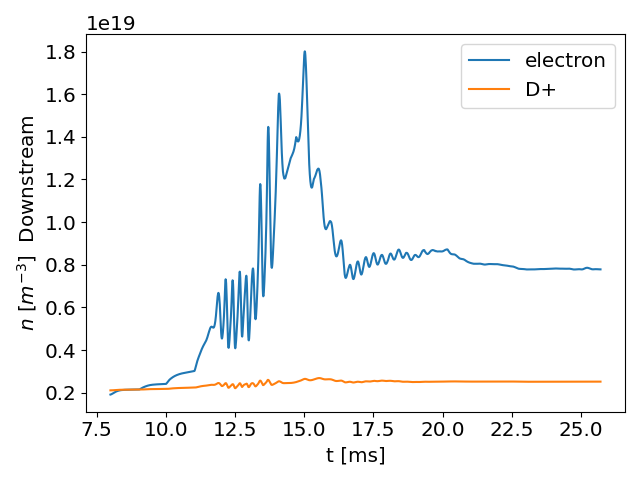}
    }\\
    \vspace{-1cm}
    \subfloat[\label{fig:trace_ardensity_down} ]{
    \includegraphics[width=0.4\textwidth]{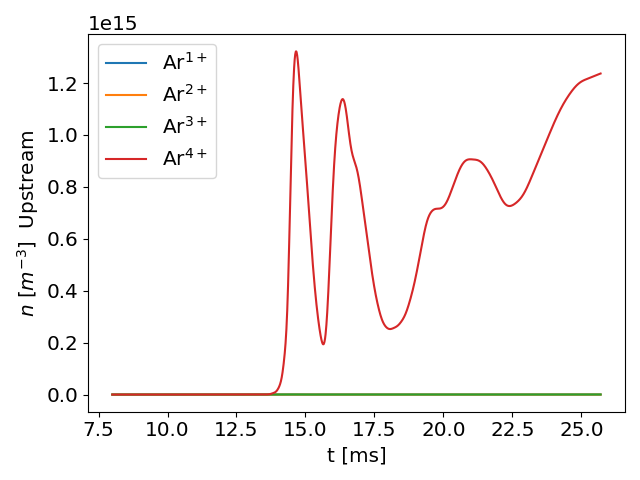}
    }\\
    \vspace{-1cm}
    \subfloat[\label{fig:trace_ardensity_down} ]{
    \includegraphics[width=0.4\textwidth]{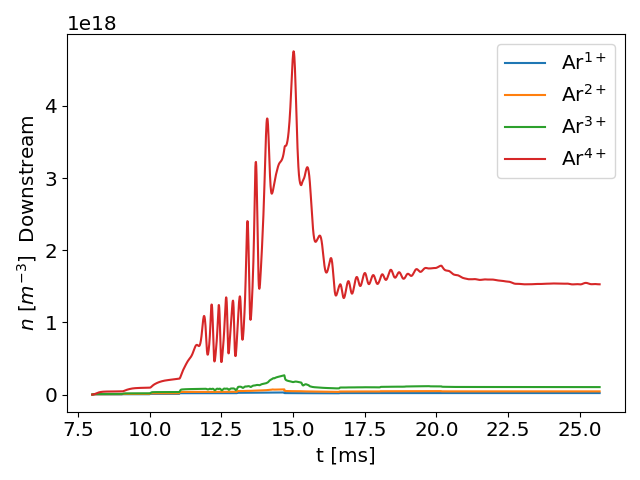}
    }
    \caption{
    \label{fig:density_trace}
    Time traces of the electron and deuterium density upstream (a) and downstream (b) and of the argon density upstream (c) and downstream (d) at the radial center of the domain from the Gkeyll simulation  in Sec.~\ref{sec:high_frac}.
    }
\end{figure}

\begin{figure}
    \subfloat[\label{fig:trace_temp} ]{
    \includegraphics[width=0.4\textwidth]{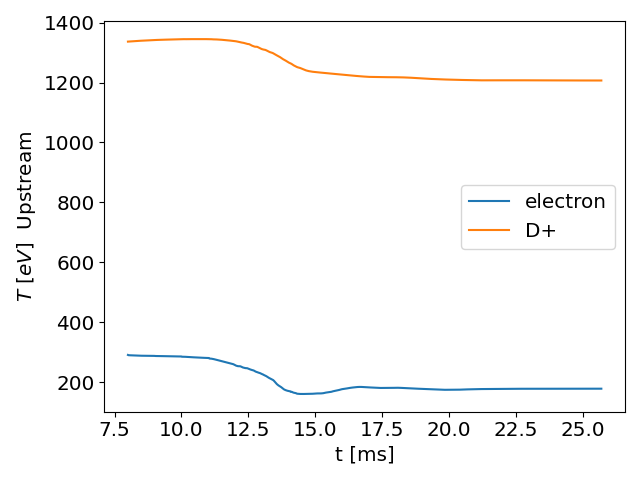}
    }\\
    \vspace{-1cm}
    \subfloat[\label{fig:trace_temp_down} ]{
    \includegraphics[width=0.4\textwidth]{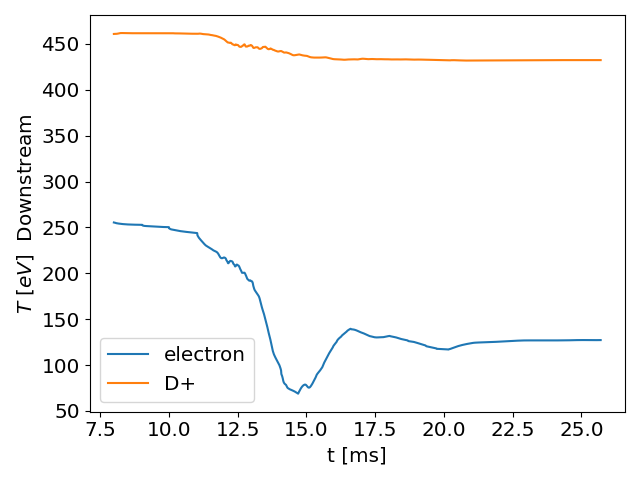}
    }\\
    \vspace{-1cm}
    \subfloat[\label{fig:trace_artemp_down} ]{
    \includegraphics[width=0.4\textwidth]{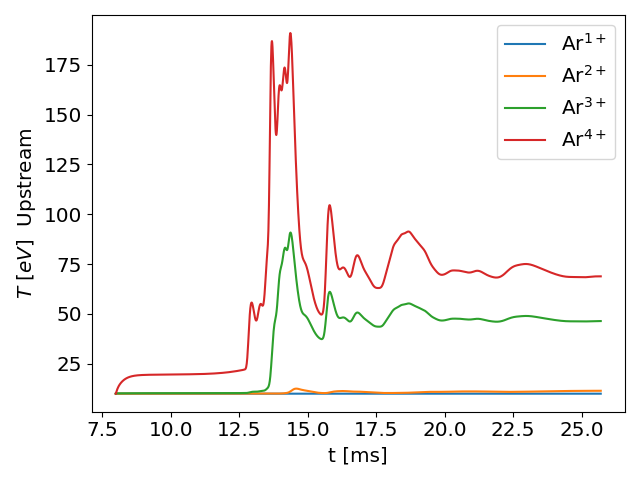}
    }\\
    \vspace{-1cm}
    \subfloat[\label{fig:trace_artemp_down} ]{
    \includegraphics[width=0.4\textwidth]{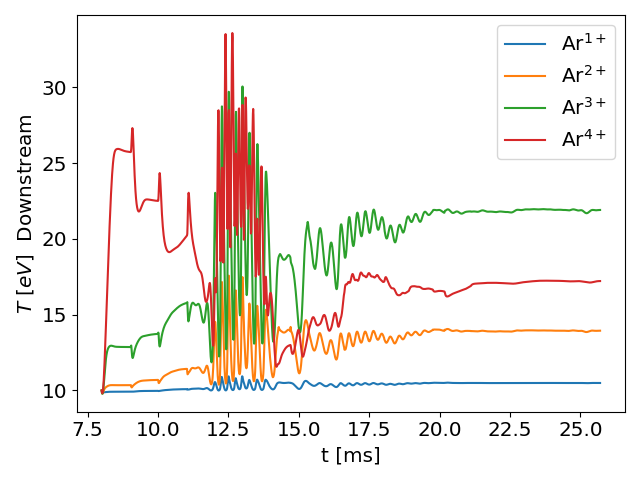}
    }
    \caption{
    \label{fig:temp_trace}
    Time traces of the electron and deuterium temperature upstream (a) and downstream (b) and of the argon temperature upstream (c) and downstream (d) at the radial center of the domain from the Gkeyll simulation  in Sec.~\ref{sec:high_frac}.
    }
\end{figure}

As noted in the main text, the high radiation case is not completely converged. In this appendix we include time traces from the Gkeyll simulation in Sec.~\ref{sec:high_frac} of the temperature and density of all species at the radial both upstream (at the midplane) and downstream (at the divertor plate). Densities are plotted in Fig.~\ref{fig:density_trace} and temperatures are plotted in Fig.~\ref{fig:temp_trace}.

All quantities except the upstream density of Ar$^{4+}$ seem to have converged both upstream and downstream. The upstream density of Ar$^{4+}$ has not completely flattened out, but it is more than 100 times lower than the upstream SOLPS density and does not seem to be increasing very fast.

As noted in Appendix~\ref{sec:cost}, this simulation  was restarted from the steady state of the  simulation in Sec.~\ref{sec:plasma_only}. This is why all the time traces start from 8 ms.

\nocite{*}
\bibliography{2dsol}

\end{document}